\documentclass[journal]{IEEEtran}
\usepackage{amsmath,amssymb}
\usepackage[dvips]{graphicx}
\usepackage{amsfonts}
\usepackage[mathscr]{eucal}
\usepackage{latexsym}
\usepackage{amsthm}
\usepackage{exscale}
\usepackage[mathscr]{eucal}
\usepackage{bm}
\usepackage[dvipsnames]{color}
\usepackage{cases}
\usepackage{epsfig}
\usepackage[center,small]{caption}
\usepackage{algorithm}
\usepackage{algorithmic}
\usepackage[verbose,nospace,sort]{cite}
\usepackage{tabularx}
\usepackage{multirow}
\usepackage{multicol}
\usepackage{balance}
\usepackage{bookmark}
\usepackage{amsmath}
\usepackage{amssymb}
\usepackage{mathrsfs}
\graphicspath{{./figs/}}

\usepackage{graphicx}

\usepackage[labelformat=brace]{subcaption}

\ifCLASSINFOpdf

\else

\fi

\scrollmode

\newtheorem{Lemma}{Lemma}

\begin{document}


\title{Multi-Antenna UAV Data Harvesting: Joint Trajectory and Communication Optimization}

\author{Jingwei~Zhang,
       Yong~Zeng,~\IEEEmembership{Member,~IEEE,}
        and~Rui~Zhang,~\IEEEmembership{Fellow,~IEEE}

\thanks{
J. Zhang and R. Zhang are with the Department of Electrical and
Computer Engineering, National University of Singapore, Singapore 117583 (e-mail:~jingwei.zhang@u.nus.edu, elezhang@nus.edu.sg).}
\thanks{Y. Zeng is with the National Mobile Communications Research Laboratory, Southeast University, Nanjing 210096, China. He
is also with the Purple Mountain Laboratories, Nanjing 211111, China (e-mail: yong\_zeng@seu.edu.cn).}
}


\maketitle

\begin{abstract}
Unmanned aerial vehicle (UAV)-enabled communication is a promising technology to extend coverage and enhance throughput for traditional terrestrial wireless communication systems. In this paper, we consider a UAV-enabled wireless sensor network (WSN), where a multi-antenna UAV is dispatched to collect data from a group of sensor nodes (SNs). The objective is to maximize the minimum data collection rate from all SNs via jointly optimizing their transmission scheduling and power allocations as well as the trajectory of the UAV, subject to the practical constraints on the maximum transmit power of the SNs and the maximum speed of the UAV. The formulated optimization problem is challenging to solve as it involves non-convex constraints and discrete-value variables. To draw useful insight, we first consider the special case of the formulated problem by ignoring the UAV speed constraint and optimally solve it based on the Lagrange duality method. It is shown that for this relaxed problem, the UAV should hover above a finite number of optimal locations with different durations in general. Next, we address the general case of the formulated problem where the UAV speed constraint is considered and propose a traveling salesman problem (TSP)-based trajectory initialization, where the UAV sequentially visits the locations obtained in the relaxed problem with minimum flying time. Given this initial trajectory, we then find the corresponding transmission scheduling and power allocations of the SNs and further optimize the UAV trajectory by applying the block coordinate descent (BCD) and successive convex approximation (SCA) techniques. Finally, numerical results are provided to illustrate the spectrum and energy efficiency gains of the proposed scheme for multi-antenna UAV data harvesting, as compared to benchmark schemes.
\end{abstract}


\section{Introduction}

As an essential part of the Internet of Things (IoT), wireless sensor networks (WSNs) have found a proliferation of applications in many fields, such as surveillance and monitoring, automated and cyber-physical systems, and so on \cite{1197877}. An important function of WSNs is data harvesting from a set of distributed sensor nodes (SNs). Various techniques such as clustering, multihop data relaying, and in-network data aggregation have been proposed to improve the data collection efficiency and reduce the energy consumption of the SNs in WSNs \cite{rawat2014wireless}. However, due to the static network infrastructure and the short transmission range of low-power SNs, it is generally difficult for WSNs to sustain a high throughput in the long term \cite{wang2010networked}.

On the other hand, thanks to various advantages such as flexible and on-demand deployment and high probability of having line-of-sight (LoS) communication links with the ground terminals, unmanned aerial vehicle (UAV) has been envisioned as a promising technique for the future wireless communication systems to support massive IoT devices \cite{zeng2019accessing}. There are various UAV applications in wireless communication systems such as UAV-aided ubiquitous coverage, UAV-aided relaying, and UAV-aided information dissemination and data collection. In particular, by leveraging the high and controllable mobility of UAVs, UAV-mounted access point (AP) is regarded as a promising new solution to collect data from widely spread SNs in a geographically large area. By properly designing the UAV trajectory to move closer to the SNs, not only the coverage and throughput performance of the WSN can be significantly improved as compared to the traditional fixed APs on the ground, but also the energy consumption of SNs can be considerably reduced, thus prolonging the lifetime of WSNs. However, UAV-enabled data harvesting in WSNs also faces new design challenges. Firstly, the trajectory of the UAV AP needs to be jointly designed with the conventional transmission scheduling and power control of the SNs to maximize the spectrum efficiency for data collection \cite{zeng2016throughput}. Secondly, besides the limited energy of SNs, UAVs are usually battery-powered with limited endurance, which needs to be taken into account in the above joint design to minimize the UAV's propulsion energy consumption \cite{7888557}, \cite{zeng2019energy}.

In recent years, significant research efforts have been devoted to designing and optimizing the performance of UAV-enabled data harvesting systems. To reduce the energy consumption of SNs, their wake-up scheduling was jointly designed with the UAV's trajectory in \cite{8119562} to minimize the maximum energy consumption of all SNs, subject to the data rate requirement for each SN. In \cite{choi2019analysis}, the performance of data harvesting from massive IoT devices to a group of UAVs was analyzed and evaluated in terms of coverage and rate. In \cite{8698468}, the UAV trajectory was optimized in three-dimensional (3D) space under the angle-dependent Rician fading channel model between the UAV and its served SNs on the ground. The projection-based compressive data gathering was investigated in \cite{8515012} for energy-efficient UAV data collection from distributed SNs. In \cite{rhc}, a low-complexity trajectory design algorithm was proposed based on the principle of receding horizon. However, the above prior works considered the scenarios that the UAV is equipped with a single antenna and thus only one SN can be scheduled for transmission at each time instant. As a result, the UAV usually needs to move sufficiently close to each SN for data collection, which leads to not only high energy consumption of the UAV, but also limited throughput for data collection from SNs.

To tackle the above issue, in this paper we propose the use of multiple antennas at the UAV for simultaneously harvesting data from multiple SNs by exploiting the multi-antenna beamforming and spatial multiplexing gains. This will help improve the spectrum efficiency for data collection as compared to the conventional single-antenna UAVs, and also reduce the UAV flying distance and hovering time and hence its propulsion energy consumption. In \cite{6127578}, the authors reported measurement results that show significant performance gains of an airborne multiple-input multiple-output (MIMO) system over the single-input single-output (SISO) system in terms of coverage and throughput. In \cite{6214709}, the heading direction of a multi-antenna UAV was optimized for maximizing the sum-rate from a set of ground nodes in their uplink communication. An antenna array composed of multiple single-antenna UAVs was proposed in \cite{8469055} to provide services to ground users in a collaborative manner, where UAVs are only allowed to communicate with ground users at a set of hovering locations, but no communication occurs when UAVs are moving. In \cite{8485372}, a virtual MIMO link was formed where a multi-antenna UAV was deployed to serve a cluster of IoT devices with the aim of maximizing the data collection efficiency. However, the UAV was assumed to follow a circular trajectory and as such, the flexible mobility of the UAV was not fully exploited for performance optimization. To our best knowledge, the problem of jointly optimizing SNs' transmission scheduling and multi-antenna UAV's trajectory has not been rigorously studied in the literature yet, to fully exploit the spatial multiplexing gain as well as the UAV mobility gain. This thus motivates our current work to study this problem from an optimization perspective.

Specifically, in this paper, we consider a UAV-enabled WSN, where a group of single-antenna SNs send their independently sensed data to a multi-antenna UAV in a periodic manner. To eliminate the inter-user interference among transmitting SNs at each time instant, zero-forcing (ZF)-based receive beamforming is adopted at the UAV. The main contributions of this paper are summarized below.

\begin{itemize}
\item First, we formulate an optimization problem to maximize the minimum data collection rate from all SNs via jointly optimizing the UAV trajectory as well as the transmission scheduling and power allocations of the SNs, subject to the practical constraints on the maximum transmit power of the SNs and the maximum flying speed of the UAV. The formulated problem is a mixed-integer non-convex optimization problem, which is difficult to be optimally solved in general.
\item Next, to tackle this problem and draw useful insight, we consider the special case of this problem by ignoring the UAV's maximum speed constraint. For this relaxed problem, it is shown that the strong duality holds and thus it can be optimally solved by employing the Lagrange duality method. It is revealed that the UAV should hover at a finite number of locations with optimal hovering durations, and this solution becomes asymptotically optimal when the UAV's flight duration and/or maximum speed becomes sufficiently large such that its flying time is negligible as compared to its hovering time.
\item Then, we address the minimum-rate maximization problem for the general case with the UAV speed constraint considered. First, we propose an efficient initial trajectory for the UAV by solving an equivalent traveling salesman problem (TSP), i.e., minimizing the UAV's flying time to sequentially visit all those hovering locations obtained via solving the previous relaxed problem. Given the initial trajectory, we then propose a suboptimal solution to the general problem by applying the block coordinate descent (BCD) and successive convex approximation (SCA) techniques, through iteratively updating the transmission scheduling/power allocations of the SNs and the trajectory of the UAV.
\item Finally, extensive simulation results are provided to evaluate the performance of the proposed design for multi-antenna UAV-enabled data harvesting. It is shown that, as compared to the maximal ratio combining (MRC) scheme (that exploits the beamforming gain only) and the single-antenna scheme, the proposed scheme requires fewer hovering locations and significantly improves the max-min rate. Furthermore, the UAV flight time required by the proposed scheme to meet the same throughput requirement of all SNs is also drastically reduced, thus greatly saving the UAV energy consumption.
\end{itemize}

The remainder of this paper is organized as follows. Section \ref{syst_mod} introduces the system model and problem formulation. Section \ref{optsol} considers the special case without UAV speed constraint and obtains the optimal solution to the corresponding relaxed problem. Section \ref{withspeed} proposes an efficient suboptimal solution to the general problem with UAV speed constraint included. Finally, numerical results are given in Section \ref{simul}, followed by the conclusions in Section \ref{conc}.

In this paper, scalars and vectors are represented by italic letters and boldface lower-case letters, respectively. $\textbf{I}$ and $\textbf{0}$ denote an identity matrix and an all-zero matrix, respectively, with appropriate dimensions. For a square matrix $\textbf{S}$, $[\textbf{S}]_{k,k}$ denotes its $k$th diagonal element. For a matrix $\textbf{M}$ of arbitrary size, $\textbf{M}^T$ and $\textbf{M}^H$ denote its transpose and conjugate transpose, respectively. $\mathbb{E}[\cdot]$ denotes the statistical expectation. $||\cdot||$ denotes the $l_2$ norm. The distribution of a circularly symmetric complex Gaussian (CSCG) random vector with mean $\boldsymbol{x}$ and covariance matrix $\boldsymbol{\Sigma}$ is denoted by $\mathcal{CN}(\boldsymbol{x},\boldsymbol{\Sigma})$, and $\sim$ stands for ``distributed as".

\section{System Model and Problem Formulation}
\label{syst_mod}
\subsection{System Model}

\begin{figure}
\vspace{-0.3cm}
\hspace*{-0.4cm}\includegraphics[width=8in]{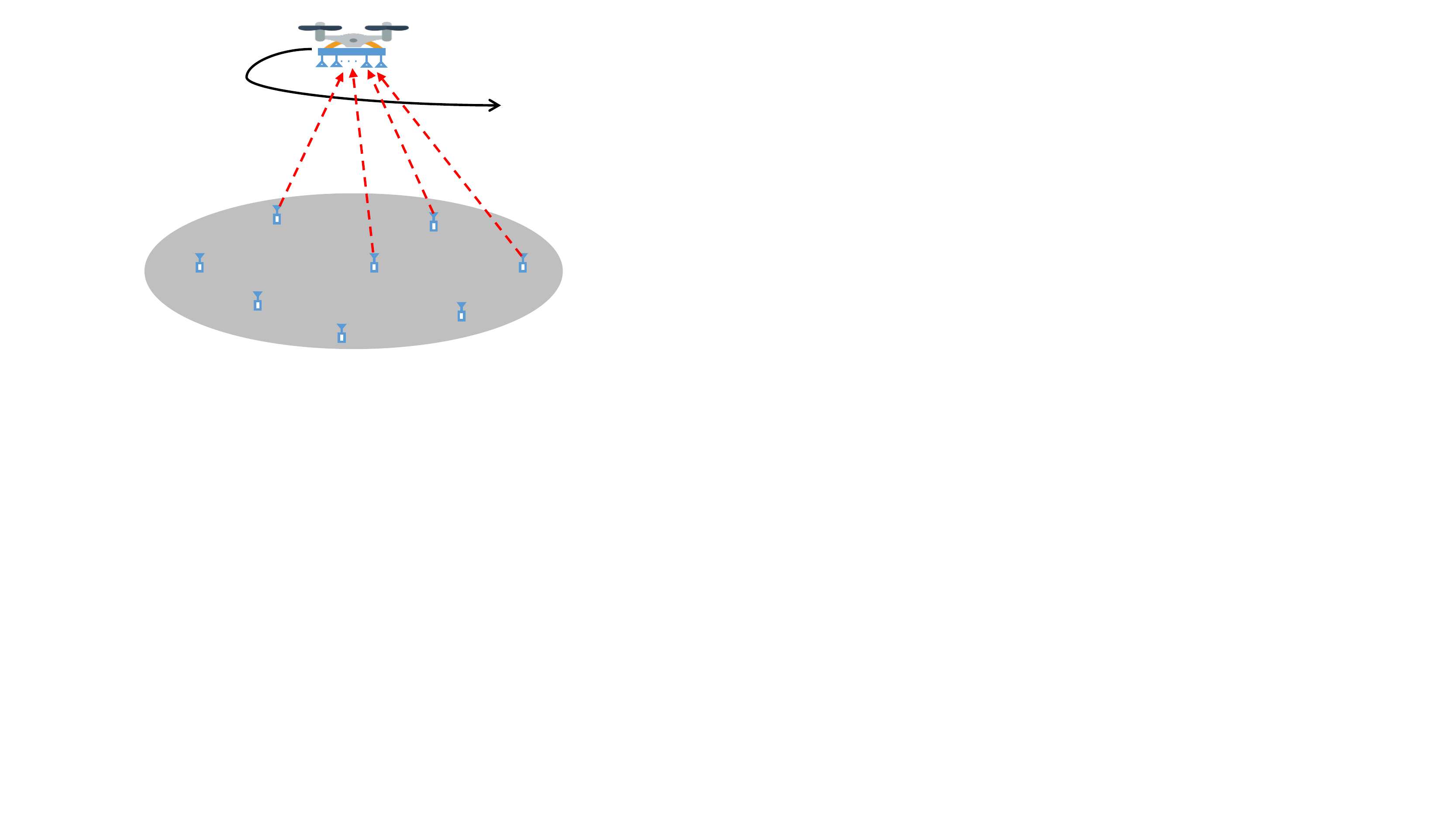}
\vspace{-7cm}
\caption{\label{syst} Multi-antenna UAV-enabled WSN. }
\vspace{-0.5cm}
\end{figure}

As shown in Fig. \ref{syst}, we consider a UAV-enabled WSN, where a UAV equipped with $M$ antennas is dispatched to collect data from $K$ distributed single-antenna SNs. The set of SNs is denoted by $\mathcal{K}=\{1,\cdots,K\}$. We consider the uplink transmission from SNs to the UAV for data collection, where the results can be similarly extended to the downlink transmission as well.

We consider a 3D cartesian coordinate system, where the location of SN $k \in \mathcal{K}$ is denoted as $[\textbf{d}_k^T,0]^T \in \mathbb{R}^{3 \times 1}$ with $\textbf{d}_k=[x_k,y_k]^T$ denoting the horizontal coordinate. For ease of exposition, the time discretization technique is applied, where the time horizon $T$ for each periodic data collection operation of the UAV is divided into $N$ time slots, each with equal length $\delta$, i.e., $t_n=n \delta$, $n=1,\cdots,N$ \cite{zeng2019accessing}. As such, the UAV trajectory is approximated by a finite number of line segments with endpoints $[\textbf{q}[n]^T,z[n]]^T \in \mathbb{R}^{3 \times 1}$, with $\textbf{q}[n]=[x[n],y[n]]^T$ and $z[n]$ representing the horizontal and vertical coordinates, respectively. Thus, the time-dependent distance between the UAV and SN $k$ is expressed as
\begin{align}
d_k[n]=\sqrt{z[n]^2+||\textbf{q}[n]-\textbf{d}_k||^2}, \  \ k \in \mathcal{K}.
\end{align}

Let the baseband equivalent complex channel between SN $k$ and the UAV be modelled as
\begin{align}
\label{chan0}
\textbf{h}_k[n]=\sqrt{\beta_k[n]}\textbf{g}_k[n],
\end{align}
where $\beta_k[n]$ denotes the large-scale channel power gain over time due to the distance-dependent path loss and shadowing, and $\textbf{g}_k[n]$ models the small-scale channel fading. Specifically, $\beta_k[n]$ is modelled as
\begin{align}
\label{chan1}
\beta_k[n]=\beta_0 d_k^{-\alpha}[n],
\end{align}
where $\beta_0$ represents the channel power gain at the reference distance of $d_0=1$ m, and $\alpha\geq 2$ is the path loss exponent. Furthermore, $\textbf{g}_k[n]$ is modelled as the Rician fading with $\mathbb{E}[||\textbf{g}_k[n]||^2]=1$ and
\begin{align}
\label{chan2}
\textbf{g}_k[n]=\sqrt{\frac{G}{G+1}}\bar{\textbf{g}}_k[n]+\sqrt{\frac{1}{G+1}}\tilde{\textbf{g}}_k[n],
\end{align}
where $G$ is the Rician factor;
$\bar{\textbf{g}}_k[n]=[e^{j \theta_{k,1}}[n],\cdots, e^{j\theta_{k,M}}[n]]^T$ denotes the LoS channel component with $\theta_{k,m}[n]$ representing the phase of the~LoS path between SN $k$ and the $m$th antenna of the UAV; $\tilde{\textbf{g}}_k[n]$ $\sim$ $\mathcal{CN}(\textbf{0},\textbf{I}_M)$ denotes the Rayleigh fading channel component.

Moreover, define a binary variable $a_k[n]$, which indicates that SN $k$ is scheduled for transmission to the UAV in time slot $n$ if $a_k[n]=1$ and otherwise if $a_k[n]=0$. We then have the following constraints,
\begin{align}
a_k[n]\in \{0,1\}, \ \ k \in \mathcal{K}, \forall n.
\end{align}
Define $\mathcal{K}_n=\{k\in\mathcal{K}: a_k[n]=1\}$ as the set of transmitting SNs in time slot $n$, and $K_n=|\mathcal{K}_n|$. Then the corresponding channel vectors are denoted by $\textbf{H}[n]=[\textbf{h}_{\mathcal{K}_n(1)}[n],\cdots,\textbf{h}_{\mathcal{K}_n(K_n)}[n]]$. By denoting the transmit power of SN $k$ in time slot $n$ as $ p_k[n]$, the received signal at the UAV can be expressed as
\begin{align}
\textbf{y}[n]=\sum_{k \in \mathcal{K}_n} \sqrt{p_k[n]}\textbf{h}_k[n]s_k[n]+\textbf{z}[n] \nonumber \\
=\textbf{H}[n]\textbf{P}[n]+\textbf{z}[n],~~~~~~~~~~~~~~~~
\end{align}
where $s_k[n]\sim\mathcal{CN}(0,1)$ denotes the signal sent by SN $k\in\mathcal{K}_n$ in time slot $n$, and $\textbf{z}[n]\sim\mathcal{CN}(\textbf{0},\sigma^2\textbf{I}_{M})$ represents the additive white Gaussian noise (AWGN) at the UAV receiver with $\sigma^2$ denoting the noise power, and $\textbf{P}[n]=[a_{\mathcal{K}_n(1)}[n]\sqrt{p_{\mathcal{K}_n(1)}[n]}s_{\mathcal{K}_n(1)}[n],\cdots,$ $a_{\mathcal{K}_n(K_n)}[n]\sqrt{p_{\mathcal{K}_n(K_n)}[n]}s_{\mathcal{K}_n(K_n)}[n]]^T$.

With linear receive beamforming applied at the UAV, the processed signal is given by
\begin{align}
\tilde{\textbf{s}}[n]=\textbf{W}[n]^H \textbf{y}[n],
\end{align}
where $\textbf{W}[n]=[\textbf{w}_{\mathcal{K}_n(1)}[n],\cdots,\textbf{w}_{\mathcal{K}_n(K_n)}[n]]$, with $\textbf{w}_k[n] \in \mathcal{C}^{M \times 1}$ denoting the beamforming vector for extracting the signal of SN $k$ and $||\textbf{w}_k[n]||=1$, $k \in \mathcal{K}_n$.

In particular, assuming the practical ZF beamforming, the inter-user interference among SNs can be completely eliminated, i.e., $\textbf{w}_k[n]^H \textbf{h}_l[n]=0$, $l \neq k$, $k,l \in \mathcal{K}_n$. Let
\begin{align}
\bar{\textbf{W}}[n]\triangleq[\bar{\textbf{w}}_1[n],\cdots,\bar{\textbf{w}}_{K_n}[n]]
=\textbf{H}[n]\left(\textbf{H}[n]^H \textbf{H}[n] \right)^{-1}.
\end{align}
Then the ZF beamforming vector for SN $k\in \mathcal{K}_n$ is given by
\begin{align}
\textbf{w}_k[n]=\frac{\bar{\textbf{w}}_k[n]}{||\bar{\textbf{w}}_k[n]||}, \ \ k \in \mathcal{K}_n, \forall n.
\end{align}

As a result, the receive signal-to-noise ratio (SNR) for decoding the signal from SN $k$ in time slot $n$ is given by
\begin{align}
\gamma_k[n]=\frac{ p_k[n]|\textbf{w}_k[n]^H\textbf{h}_k[n]|^2}{\sigma^2}, \nonumber ~~~~~~~~~~~~~~~~~\\
=\frac{p_k[n]}{ \left[ (\textbf{H}[n]^H\textbf{H}[n])^{-1} \right]_{k,k}\sigma^2}, \ \   k \in \mathcal{K}_n,\forall n.\!\!\!\!
\end{align}
By averaging over the random small-scale channel fading, the achievable rate of SN $k$ over the $n$th time slot in bits/second/Hertz (bps/Hz) is given by
\begin{align}
\label{rate}
R_k[n]=a_k[n]\mathbb{E}\left[\log_2\left(1+\frac{p_k[n]|\textbf{w}_k[n]^H \textbf{h}_k[n]|^2}{\sigma^2} \right) \right] \nonumber \\
=a_k[n]\mathbb{E}\left[\log_2\left(1+\frac{p_k[n]}{ \left[ (\textbf{H}[n]^H\textbf{H}[n])^{-1} \right]_{k,k}\sigma^2} \right)   \right], \nonumber  \!\!\!\!\!\!\!\!\!\! \\
k \in \mathcal{K}_n, \forall n.
\end{align}
The average achievable rate for SN $k$ over the entire time horizon is then obtained as
\begin{align}
\bar{R}_k=\frac{1}{N} \sum_{n=1}^N R_k[n] , \ \ k \in \mathcal{K}.
\end{align}

Note that $R_k[n]$ in general does not admit a closed-form expression due to the difficulty in handling the expectation operation in \eqref{rate}. Thus, in the following, we derive a lower bound of $R_k[n]$ based on the results in \cite{liu2019comp}. Specifically, since both the LoS component and Rayleigh fading component in the Rician channel model are independent over different SNs within each time slot, a tight lower bound of $R_k[n]$ can be derived as
\begin{align}
\label{R1}
\!\!\!\!\!R_k[n] \geq a_k[n] \log_2\left(1+\frac{p_k[n]}{\mathbb{E}\left[  \left[ (\textbf{H}[n]^H\textbf{H}[n])^{-1} \right]_{k,k}\right]\sigma^2} \right)~~~~~\!\!\!\!\!\!\!\!\!\!\!\! \nonumber \\
= a_k[n] \log_2\left(1+\frac{p_k[n] \beta_0 d_k[n]^{-\alpha}}{\frac{\sigma^2}{M-K_n}} \right) \nonumber~~~~~~~~~~~~~~~~~~\!\!\!\!\!\!\!\!\!\! \\
\!\!=a_k[n] \log_2\left(1+ \frac{(M-K_n)p_k[n]\gamma_0}{(z[n]^2+||\textbf{q}[n]-\textbf{d}_k||^2)^{\alpha/2}}  \right),
\end{align}
where $\gamma_0 \triangleq \beta_0/\sigma^2$. It should be noted that the above rate lower bound of $R_k[n]$ only applies for $K_n\geq 2$ for the feasibility of ZF beamforming assumed. While if $K_n=1$, we can simply employ the optimal MRC-based receive beamforming at the UAV such that the achievable rate can also be approximated similarly as \eqref{R1} with the term $(M-K_n)$ replaced by $M$. Hence, for consistency, we combine the above two cases and approximate $R_k[n]$ as
\begin{align}
\label{R2}
\!\!r_k[n]\triangleq a_k[n]\log_2\left(1+\frac{\kappa_np_k[n]\gamma_0}{\left(z[n]^2+||\textbf{q}[n]-\textbf{d}_k||^2 \right)^{\alpha/2}} \right),
\end{align}
where
\begin{align}
\kappa_n =\left\{
               \begin{array}{ll}
                 M-K_n, & K_n \geq 2, \\
                 M, &     K_n=1.
               \end{array}
             \right.
\end{align}
Then, the average achievable rate is approximated as $\bar{r}_k \triangleq \frac{1}{N} \sum_{n=1}^N r_k[n]$, $k \in \mathcal{K}$. Numerical results in Section \ref{simul} will show that the above rate approximation achieves good accuracy, especially for large $M$.

From \eqref{R2}, it is observed that the achievable rate of each transmitting SN in each time slot (i.e., $a_k[n]=1$) depends on not only its distance from the UAV, but also the spatial multiplexing gain of the UAV (i.e., $\kappa_n$). Moreover, it is noted that to achieve the maximum transmission rate, the UAV should always fly at the minimum possible altitude, i.e., $z[n]=H_{\mathrm{min}}$, $\forall n$, to minimize the distances to all SNs, where $H_{\mathrm{min}}$ corresponds to the minimum altitude allowed to ensure safety in practice.

\subsection{Problem Formulation}

Our objective is to maximize the minimum average rate from all SNs by jointly optimizing the UAV trajectory $\{\textbf{q}[n]\}$, SNs' transmission scheduling $\{a_k[n]\}$ and power allocations $\{p_k[n]\}$. By defining $r\triangleq \min_{ k \in \mathcal{K}} \bar{r}_k$, this optimization problem is formulated as
\begin{subequations}
\begin{align}
\mathrm{(P1)}~~\max_{\substack{\{\textbf{q}[n]\},\{a_k[n]\},\\ \{p_k[n]\},r  }}  r~~~~~~~~~~~~~~~~~  ~~  \nonumber \\
\label{p1001}
\mathrm{s.t.}~~ \frac{1}{N} \sum_{n=1}^N r_k[n]\geq r, \ \ k \in \mathcal{K}, ~~~~~~~~~~~~~~~~~\\
\label{p1002}
\sum_{k=1}^K a_k[n] \leq M, \ \ \forall n,~~~~~~~~~~~~~~~~~~~~~~~
\end{align}
\begin{align}
\label{p1003}
a_k[n] \in \{0,1\}, \ \ k \in \mathcal{K}, \forall n, ~~~~~~~~~~~~~~~\\
\label{p1004}
\frac{1}{N}\sum_{n=1}^N p_k[n] \leq \bar{P}, \ \ k \in \mathcal{K},~~~~~~~~~~~~~~~~ \\
\label{p1005}
p_k[n] \geq 0, \ \ k \in \mathcal{K},  \forall n,~~~~~~~~~~~~~~~~~~~~ \\
\label{p1006}
||\textbf{q}[n+1]-\textbf{q}[n]|| \leq V_{\mathrm{h}},  \ \ n=1,\cdots,N-1, \!\!\!\!\!\!\!\!\!\!\!\! \\
\label{p1007}
\textbf{q}[1]=\textbf{q}_{\mathrm{I}}, \ \ \textbf{q}[N]=\textbf{q}_{\mathrm{F}},~~~~~~~~~~~~~~~~~~~~\!
\end{align}
\end{subequations}
where $\bar{P}$ is the average power limit at each SN, and $V_{\mathrm{h}}\triangleq v_{\mathrm{h}} \delta$ with $v_{\mathrm{h}}$ representing the maximum horizontal speed of the UAV, $\textbf{q}_{\mathrm{I}}$ and $\textbf{q}_{\mathrm{F}}$ are the initial and final locations of the UAV, respectively.

Note that problem (P1) is challenging to solve due to the following reseasons. Firstly, $r_k[n]$ in constraint \eqref{p1001} is not jointly concave with respect to the optimization variables $\{\textbf{q}[n]\}$, $\{a_k[n]\}$ and $\{p_k[n]\}$. Secondly, even with fixed trajectory $\{\textbf{q}[n]\}$, problem (P1) is still a mixed-integer non-linear programming as the binary variable $a_k[n]$ is in general coupled with $\kappa_n$ and $p_k[n]$ in $r_k[n]$ given by \eqref{R2}. Consequently, problem (P1) is a mixed-integer non-convex optimization problem, which is difficult to solve in general. To tackle this problem, we first consider the special case of (P1) by ignoring the UAV maximum speed constraint in \eqref{p1006} as well as the initial/final location constraints in \eqref{p1007}. The relaxed problem of (P1) is thus given by
\begin{subequations}
\begin{align}
\mathrm{(P2)}~~\max_{\substack{\{\textbf{q}[n]\},\{a_k[n]\},\\ \{p_k[n]\},r  }}  r~~~~~~~~~~~  ~~  \nonumber \\
\mathrm{s.t.}~~\eqref{p1001}-\eqref{p1005}. ~~~~~~~~~~~~~~~ \nonumber
\end{align}
\end{subequations}
In the following, we first find the optimal solution to (P2) in Section \ref{optsol}. Then based on the optimal solution obtained, we propose an efficient solution to solve the general problem (P1) sub-optimally in Section \ref{withspeed}.

\section{Optimal Solution to (P2)}
\label{optsol}

In this section, we solve the relaxed problem (P2), which is still challenging due to the non-convex constraints and binary variables. Fortunately, it can be verified that (P2) satisfies the so-called time-sharing condition in \cite{yu2006dual}, so that the strong duality holds between (P2) and its Lagrange duality problem. Therefore, (P2) can be optimally solved by the Lagrange duality method.

Specifically, the partial Lagrangian of (P2) is given by
\begin{align}
\!\!\!\!\!\!\!\!\!\!\!\!\!\!\!\!\!\!\!\!\!\!\!\!\!\!\!\!\!\!\!\!\!\!\!\!\!\mathcal{L}_1( \{\textbf{q}[n]\},\{p_k[n]\},\{a_k[n]\},r,\boldsymbol{\lambda},\boldsymbol{\mu})= \nonumber~~~~~~~~~~~~~~~~ \\
\!\!\!\!\!\!\!\!\!\!\!\!\!\!\!\!\!\!\left(1-\sum_{k=1}^K\lambda_k\right)r
+\sum_{k=1}^K \lambda_k\bar{r}_k+\sum_{k=1}^K\mu_k\left(N \bar{P}-  \sum_{n=1}^{N }p_k[n]\right),\!\!\!\!\!\!\!\!\!\!\!\!\!\!\!  \nonumber \\
\end{align}
where $\boldsymbol{\lambda}\triangleq \{\lambda_k\}$ and $\boldsymbol{\mu}\triangleq\{\mu_k\}$ are the dual variables associated with the constraints in \eqref{p1001} and \eqref{p1004}, respectively. The Lagrange dual function of (P2) is then given by
\begin{align}
\label{gg}
 \!\!\!\!\!\!g_1(\boldsymbol{\lambda},\boldsymbol{\mu})=\nonumber ~~~~~~~~~~~~~~~~~~~~~~~~~~~~~~~~~~~~~~~~~~~~~~~~~\\
\!\!\!\left\{
              \begin{aligned}
              \max_{\substack{\{\textbf{q}[n]\},\{p_k[n]\},\\ \{a_k[n]\},r}} \mathcal{L}_1\left(\{\textbf{q}[n]\},\{p_k[n]\},\{a_k[n]\},r,\boldsymbol{\lambda},\boldsymbol{\mu}\right) \\
              \mathrm{s.t.}~~~~~\eqref{p1002},~\eqref{p1003},~\eqref{p1005}.~~~~~~~~~~~~~~~~~~~
                            \end{aligned}
            \right.\!\!\!\!\!\!\!\!
\end{align}

For $g_1(\boldsymbol{\lambda},\boldsymbol{\mu})$ in \eqref{gg} to be bounded, we should have $1-\sum_{k=1}^K\lambda_k=0$. Therefore, the dual problem of (P2) is given by
\begin{subequations}
\begin{align}
\mathrm{(D2)}~~ \min_{\boldsymbol{\lambda},\boldsymbol{\mu}} ~~~g_1(\boldsymbol{\lambda},\boldsymbol{\mu})  ~~~~~~~~~~~~~~\nonumber \\
\label{d21001}
\mathrm{s.t.}~~~1-\sum_{k=1}^K\lambda_k=0,~~~~~~~~\!~~~~~ \\
\label{d21002}
\lambda_k\geq 0, \ \ \forall k \in \mathcal{K},~~~\!~~~~~~~ \\
\label{d21003}
\mu_k \geq 0, \ \ \forall k \in \mathcal{K}. ~~~~~~~~~~~\!\!
\end{align}
\end{subequations}
Since the strong duality holds between (P2) and (D2), we can solve (P2) by equivalently solving (D2). Let the feasible set of $\boldsymbol{\lambda}$ and $\boldsymbol{\mu}$ characterized by the constraints in \eqref{d21001}-\eqref{d21003} be denoted as $\mathcal{X}_1$. In the following, we first obtain $g_1(\boldsymbol{\lambda},\boldsymbol{\mu})$ by solving problem \eqref{gg} under any given $(\boldsymbol{\lambda},\boldsymbol{\mu}) \in \mathcal{X}_1$, and then solve (D2) to find the optimal  $(\boldsymbol{\lambda},\boldsymbol{\mu})$ to minimize $g_1(\boldsymbol{\lambda},\boldsymbol{\mu})$, and finally construct the optimal solution to (P2).

\subsection{Obtaining $g_1($\texorpdfstring{$\boldsymbol{\lambda}$}{}$,~\!\!$\texorpdfstring{$\boldsymbol{\mu}$}{}$)$ by solving problem \eqref{gg} for given $($\texorpdfstring{$\boldsymbol{\lambda}$}{}$,~\!\!$\texorpdfstring{$\boldsymbol{\mu}$}{}$)$ $\in \mathcal{X}_1$}

For any given $(\boldsymbol{\lambda},\boldsymbol{\mu})\in \mathcal{X}_1$, problem \eqref{gg} is still a non-convex optimization problem. In order to solve this problem, we first focus on the case with given SNs' scheduling $\{a_k[n]\}$ and trajectory $\{\textbf{q}[n]\}$ to optimize the power allocations $\{p_k[n]\}$ only, which is reformulated as
\begin{align}
\label{water}
\max_{\{p_k[n]\}}~~ \sum_{k=1}^K \lambda_k\bar{r}_k+\sum_{k=1}^K\mu_k\left(N \bar{P}- \sum_{n=1}^{N }p_k[n]\right)
\end{align}
\vspace{-0.5cm}
\addtocounter{equation}{-1}
\begin{subequations}
\begin{align}
\mathrm{s.t.}~~~p_k[n] \geq 0, \ \ k \in \mathcal{K}, \forall n. ~~~~~~~~~\!\!~~~~~~
\end{align}
\end{subequations}

As it can be verified that the objective of problem \eqref{water} is a concave function with respect to $p_k[n] $, problem \eqref{water} is a convex optimization problem. Then by applying the standard Lagrange duality method and the Karush-Kuhn-Tucker (KKT) conditions, it can be shown that the optimal solution to \eqref{water} follows the classic water-filling (WF) structure given by
\begin{align}
\label{waterp}
p_k^*[n]=\left[ \frac{\lambda_ka_k[n]}{N\mu_k \log(2)}-\frac{d_k[n]^{\alpha}}{\kappa_n\gamma_0}\right]^+, \nonumber ~~~~~~ \\
\stackrel{(a)}{=}a_k[n] \left[ \frac{\lambda_k}{N\mu_k \log(2)}-\frac{d_k[n]^{\alpha}}{\kappa_n\gamma_0}\right]^+,\!
\end{align}
where $(a)$ holds since $a_k[n]\in \{0,1\}$, and $[b]^+=\max\{b,0\}$. As a result, the achievable rate is expressed as
\begin{align}
\label{waterr}
r^*_k[n]=a_k[n]\left[\log_2\left(\frac{\lambda_ka_k[n]\kappa_n\gamma_0}{N\mu_k\log(2)d_k[n]^{\alpha}} \right) \right]^+ \nonumber ~\\
=a_k[n]\left[\log_2\left(\frac{\lambda_k\kappa_n\gamma_0}{N\mu_k\log(2)d_k[n]^{\alpha}} \right) \right]^+,
\end{align}
and thus $\bar{r}^*_k=\frac{1}{N}\sum_{n=1}^N r^*_k[n]$. With the obtained $\{p_k^*[n]\}$ and $\{r^*_k[n]\}$, problem \eqref{gg} is further recast to
\begin{align}
\label{gg1}
\max_{\{\textbf{q}[n]\},\{a_k[n]\}}\sum_{k=1}^K \lambda_k\bar{r}^*_k+\sum_{k=1}^K\mu_k\left(N \bar{P}- \sum_{n=1}^{N }p^*_k[n]\right)  \\
\mathrm{s.t.}~~~~~\eqref{p1002},~\eqref{p1003}. \nonumber~~~~~\!~~~~~~~~~~~~~~~~~~~~~~~~~~
\end{align}
It is worth noting that \eqref{gg1} consists of $N$ sub-problems, each corresponding to one time slot. As all sub-problems are identical across different time slots, we can drop the time slot index $n$ and re-express each problem as
\begin{align}
\label{gg2}
\max_{\textbf{q},\{ a_k\},\kappa}~~\sum_{k=1}^K a_kf_k(\textbf{q}, \kappa)+N\bar{P}\sum_{k=1}^K \mu_k~~~ \\
\mathrm{s.t.}~~~~~~\eqref{p1002}, \eqref{p1003},\nonumber ~~~~~~~~~~~~~~~~~~~~~
\end{align}
where
\begin{align}
\!\!\!f_k(\textbf{q},\kappa)\triangleq\lambda_k \left[\log_2\left(\frac{\lambda_k\kappa\gamma_0}{N\mu_k\log(2)d_k^{\alpha}} \right) \right]^+ ~~~~~~~\nonumber  \\
 -N\mu_k\left[ \frac{\lambda_k}{N\mu_k \log(2)}-\frac{d_k^{\alpha}}{\kappa\gamma_0}\right]^+.
\end{align}
The constant term in the objective of \eqref{gg2} has been omitted for brevity. Note that \eqref{gg2} is a mixed-integer optimization problem, which is non-convex and difficult to solve in general. In the following, we optimally solve it via a two-dimensional (2D) search.

To start with, we consider the box region $[\underline{x},\overline{x}]^T \times [\underline{y},\overline{y}]^T$, where $\underline{x} =\min_{k \in \mathcal{K}}x_k$, $\overline{x}=\max_{k \in \mathcal{K}}x_k$, $\underline{y} =\min_{k \in \mathcal{K}}y_k$, $\overline{y}=\max_{k \in \mathcal{K}}y_k$, and discretize it with a finite granularity $\Delta_g$. Note that the value of $\Delta_g$ is typically chosen to be sufficiently small to ensure certain accuracy. Next, we focus on finding $\{a_k\}$ and $\kappa$ at each discretized location $\textbf{q}$ to maximize the objective value in \eqref{gg2} and then compare them to find the maximum one.

Note that with fixed $\textbf{q}$ and $\kappa$, the values of $\{f_k(\textbf{q},\kappa)\}$ are also determined. Therefore, to maximize the objective value in \eqref{gg2} with given $\textbf{q}$ and $\kappa$, we only need to choose the $(M-\kappa)$  largest values (or choose the largest value when $\kappa=M$) among $\{f_k(\textbf{q},\kappa)\}$ and set the corresponding scheduling variables $\{a_k\}=1$. Then by comparing the objective value achieved by each possible value of $\kappa$, where $\kappa=M,M-2,M-3,\cdots,M-K_{\mathrm{max}}$ with $K_{\mathrm{max}}\triangleq \min\{M-1,K\}$, we are able to find the optimal value of $\kappa$ and the corresponding $\{a_k\}$ under the given location $\textbf{q}$. Finally, by comparing the objective values achieved at each discretized location, we can find the optimal solution to problem \eqref{gg2}, which is given by
\begin{align}
\textbf{q}^{\{ \boldsymbol{\lambda},\boldsymbol{\mu} \}}[n]=\textbf{q}^{\{ \boldsymbol{\lambda},\boldsymbol{\mu} \}},  \ \ \{a^{\{ \boldsymbol{\lambda},\boldsymbol{\mu} \}}_k[n]\}=\{a^{\{ \boldsymbol{\lambda},\boldsymbol{\mu} \}}_k\},  \nonumber \\
 \kappa^{\{ \boldsymbol{\lambda},\boldsymbol{\mu} \}}_n=\kappa^{\{ \boldsymbol{\lambda},\boldsymbol{\mu} \}}, \ \
\forall n.
\end{align}
It is worth pointing out that although the above optimal solution to \eqref{gg2} is generally non-unique, we can arbitrarily choose one of them to obtain the dual function $g_1(\boldsymbol{\lambda},\boldsymbol{\mu})$ in (D2). Note that as the optimal dual solution is generally non-unique, the solution obtained for (D2) may not be optimal for (P2) after solving the dual problem; thus, an additional step is usually needed to construct the optimal solution to (P2) based on the dual optimal solution, as will be shown in Section \ref{construct}.


\subsection{Finding optimal $ $\texorpdfstring{$\boldsymbol{\lambda}$}{}$ $ and $ $\texorpdfstring{$\boldsymbol{\mu}$}{}$ $ to solve (D2) }
\label{timeshar2}

In the following, we search over $\boldsymbol{\lambda}$ and $\boldsymbol{\mu}$ to minimize $g_1(\boldsymbol{\lambda},\boldsymbol{\mu})$ for solving (D2). Since the dual problem (D2) is always convex but non-differentiable in general, we employ the subgradient based method, such as the ellipsoid method \cite{boyd2012ee364b}, to obtain the optimal dual solution $\boldsymbol{\lambda}$ and $\boldsymbol{\mu}$, which are denoted by $\boldsymbol{\lambda}^*$ and $\boldsymbol{\mu}^*$. In each iteration, the dual variables $\boldsymbol{\lambda}$ and $\boldsymbol{\mu}$ are updated based on the objective and constraints in (D2). Specifically, the subgradients with respect to $(\boldsymbol{\lambda},\boldsymbol{\mu})$ are given by
\begin{align}
\!\Delta \lambda_k =r_k^*,~~~~~\!~~~~~~\\
\Delta \mu_k=N\bar{P}-Np_k^*,
\end{align}
where
\begin{align}
    r_k^*=a_k\left[\log_2\left(\frac{\lambda_k^*\kappa\gamma_0}{N\mu^*_k\log(2)d_k^{\alpha}} \right) \right]^+, \\
    p_k^*=a_k \left[ \frac{\lambda^*_k}{N\mu^*_k \log(2)}-\frac{d_k^{\alpha}}{\kappa \gamma_0}\right]^+. ~~~~
\end{align}

\subsection{Constructing optimal solution to (P2)}
\label{construct}
Based on the optimal dual solution $\boldsymbol{\lambda}^*$ and $\boldsymbol{\mu}^*$, it remains to obtain the optimal (primal) solution to (P2), which is denoted as $\{\textbf{q}^*[n]\}$, $\{p^*_k[n]\}$, and $\{a^*_k[n]\}$. It is worth pointing out that when the Lagrange duality method is employed to solve (P2) via (D2), the optimal solution to problem \eqref{gg} under the optimal dual solution $\boldsymbol{\lambda}^*$ and $\boldsymbol{\mu}^*$ (i.e., $\{\textbf{q}^{\boldsymbol{\lambda}^*,\boldsymbol{\mu}^*}[n]\}$, $\{p^{\boldsymbol{\lambda}^*, \boldsymbol{\mu}^*}_k[n]\}$, and $\{a_k^{\boldsymbol{\lambda}^*, \boldsymbol{\mu}^*}[n]\}$) is the optimal solution to (P2), if such a solution is feasible \cite{boyd2004convex}. On the other hand, when the obtained $\{\textbf{q}^{\boldsymbol{\lambda}^*, \boldsymbol{\mu}^*}[n]\}$, $\{p^{\boldsymbol{\lambda}^*, \boldsymbol{\mu}^*}_k[n]\}$ and $\{a_k^{\boldsymbol{\lambda}^*, \boldsymbol{\mu}^*}[n]\}$ for problem \eqref{gg} are non-unique, they may not be feasible nor optimal to problem (P2) in general. In this case, additional procedures are needed to construct the optimal solution to (P2) by applying time-sharing over these non-unique solutions.

Specifically, with the optimal dual solution $\boldsymbol{\lambda}^*$ and $\boldsymbol{\mu}^*$, suppose that problem \eqref{gg1} has a total number of $\Omega$ solutions, denoted by $\{\textbf{q}^*_{\omega}\}$, $\{p^*_{k,\omega}\}$, and $\{a^*_{k,\omega}\}$, $\omega=1,\cdots,\Omega$. The corresponding average rate is then expressed as $\bar{r}^*_{k,\omega}$. Due to the zero duality gap between (P2) and (D2), it is obvious that at the optimal solution for (P2), the UAV should choose from the $\Omega$ locations to hover above in each time slot. Since the achievable rate of SNs remains the same when the UAV stays at the same hovering location in different time slots, the optimal solution to (P2) can be constructed via allocating the flight duration $T$ over the $\Omega$ hovering locations obtained. Let $\tau_\omega$ denote the hovering duration of the UAV above the location $\textbf{q}^*_{\omega}$. The extra problem to maximize the minimum average rate of SNs is formulated as
\begin{subequations}
\begin{align}
\mathrm{(P2.1)}~~\max_{\{\tau_{\omega}\},r}~~r  ~~~~~~~~~~~~~~~\!~~~~~  \nonumber \\
\mathrm{s.t.}~~\frac{1}{T} \sum_{\omega=1}^{\Omega} \tau_{\omega} \bar{r}^*_{k,\omega}  \geq r,  \ \ k \in \mathcal{K},~ \\
\sum_{\omega=1}^{\Omega}\tau_{\omega}=T,~~~~~~~~~~~~~~~~~~~\\
\tau_{\omega} \geq 0, \ \ \omega=1,\cdots, \Omega.~~~\!~~~
\end{align}
\end{subequations}
Problem (P2.1) is a linear programming (LP), which can be efficiently solved via the standard convex optimization technique \cite{grant2014cvx}. It should be noted that after solving (P2.1), there might exist some hovering locations with $\tau_{\omega}=0$, which indicates that these hovering locations are not required for the optimal solution to (P2). Therefore, we should only choose those locations with the corresponding $\tau_{\omega}>0$ and the number of such locations is denoted as $\Omega^*$.

In summary, the details to solve (P2) are presented in Algorithm \ref{algo1}. The computational complexity of Algorithm \ref{algo1} mainly consists of three parts. The first one is the exhaustive search over the region in step 3 with complexity $O(K_{\mathrm{max}}\Delta_G)$, where $\Delta_G\triangleq (\bar{x}-\underline{x})(\bar{y}-\underline{y})/\Delta_g^2$. The second part is for updating the dual variables via the ellipsoid method in step 4 with complexity $O(K^2)$. Since the ellipsoid method takes $O(K^2)$ to converge \cite{boyd2012ee364b}, the overall complexity from step 2 to step 5 is $O((K_{\mathrm{max}}\Delta_G+K^2)K^2)$. The third part is due to step 7 for solving the LP with complexity $O(\Omega^3)$. Due to the much smaller value of $\Omega^3$ compared to $K^4$, the overall complexity of Algorithm \ref{algo1} is approximately $O((K_{\mathrm{max}}\Delta_G+K^2)K^2)$.

\begin{algorithm}
\caption{\label{algo1}Proposed Algorithm for Solving (P2).}
1:~Initialize an ellipsoid $\boldsymbol{ \epsilon}((\boldsymbol{\lambda},\boldsymbol{\mu}),\textbf{A})$ containing $(\boldsymbol{\lambda}^*,\boldsymbol{\mu}^*)$,\\
 \hspace*{0.3cm} where $(\boldsymbol{\lambda},\boldsymbol{\mu})$ is the center point of the ellipsoid, and $\textbf{A}$ is\\\hspace*{0.3cm} a positive definite matrix that characterizes its size.   \\
2:~\textbf{Repeat}\\
3:~~\!\!~~Obtain $\{\textbf{q}^{\boldsymbol{\lambda}, \boldsymbol{\mu}}\}$, $\{p_k^{\boldsymbol{\lambda}, \boldsymbol{\mu}}\}$, and $\{a_k^{\boldsymbol{\lambda}, \boldsymbol{\mu}}\}$ via 2D search over\\
\hspace*{0.55cm} the region  $[\underline{x},\overline{x}]^T \times [\underline{y},\overline{y}]^T$.\\
4:~~~~Compute the subgradients of $g_1(\boldsymbol{\lambda},\boldsymbol{\mu})$, and
update $(\boldsymbol{\lambda},\boldsymbol{\mu})$ \\
\hspace*{0.6cm}  by the ellipsoid method.~\\
5:~\textbf{Until} $\boldsymbol{\lambda}$ and $\boldsymbol{\mu}$ converge within a given accuracy.\\
6:~Set $(\boldsymbol{\lambda}^*,\boldsymbol{\mu}^*)\leftarrow (\boldsymbol{\lambda},\boldsymbol{\mu})$. \\
7:~Obtain the optimal solution to (P2) via solving (P2.1).
\end{algorithm}

Note that the above results demonstrate that the UAV should hover above a finite number of locations with optimal allocations of the hovering duration to maximize the minimum average rate of all SNs by solving problem (P2) optimally without considering the UAV speed and initial/final location constraints\footnote{Note that the initial/final UAV location constraints can be added back to (P2) without changing the above optimal solution, since if they do not belong to the optimal hovering locations, we can simply assign zero hovering time for them without loss of generality}. In practice, even with finite speed constraint for the UAV, when the given time horizon $T$ or the speed of the UAV $v_{\mathrm{h}}$ is sufficiently large such that the flying time of the UAV becomes negligible as compared its hovering time, the above solution can be shown to be asymptotically optimal to problem (P1) \cite{wu2018capacity}.

\section{Proposed Solution to (P1)}
\label{withspeed}

In this section, we consider the general problem (P1) with the UAV speed and initial/final location constraints included, i.e., the flying time of the UAV as well as the communications with the SNs during its flying cannot be ignored. Although the results obtained in Section \ref{optsol} are not directly applicable to the general case of (P1), the max-min rate achieved by solving problem (P2) can serve as an upper bound for that achievable by solving problem (P1).

In the following, we propose an efficient suboptimal solution to (P1) based on the optimal solution obtained for (P2). Specifically, we firstly design an efficient initial trajectory for the UAV based on the classic TSP, where the UAV sequentially visits the $\Omega^*$ locations obtained by solving (P2) with the minimum flying time. Then, we find the optimal transmission scheduling and power allocations of SNs with the given UAV trajectory and iteratively optimize them with the UAV trajectory by applying the BCD and SCA techniques, until all of them get converged.

\subsection{TSP-based initial trajectory}
\label{tspini}

In this subsection, we design a TSP-based initial trajectory, which minimizes the flying time of the UAV to sequentially visit those $\Omega^*$ hovering locations obtained in Section \ref{optsol}. Note that the proposed TSP-based trajectory is similar to the successive hover-and-fly trajectory proposed in \cite{zeng2019energy}, \cite{xu2018uav,xie2018throughput,zhang2018uav}.

To maximize the hovering time of the UAV above the obtained hovering locations, the UAV should always fly with the maximum speed $v_{\mathrm{h}}$ among these locations, and the flying time, or equivalently the flying distance, needs to be minimized. By including the initial/final location constraints in \eqref{p1007}, the problem of minimizing the UAV flying distance is reminiscent of the classic TSP, which can be efficiently solved via the techniques outlined in \cite{zhang2018uav}, \cite{zeng2018trajectory}. After solving the corresponding TSP, we obtain the minimum flying time required, denoted by $T_{\mathrm{tsp}}$, to visit all the hovering locations, and the permutation order $\boldsymbol{\pi}\triangleq [\pi(1),\cdots,\pi(\Omega^*) ]$, with $\pi(\omega)$ representing the index of the $\omega$th hovering location to be visited. In the following, for any given flight duration $T$, we design the initial trajectory depending on whether $T$ is greater than $T_{\mathrm{tsp}}$ or not.

\subsubsection{$ $\texorpdfstring{$T\geq T_{\mathrm{tsp}}$}{}$ $}
 \label{tsp1}
In this case, the UAV is able to reach all the hovering locations. With the permutation order $\boldsymbol{\pi}$ and flying time $T_{\mathrm{tsp}}$, the trajectory design reduces to allocating the remaining time $T-T_{\mathrm{tsp}}$ among all hovering locations, which can be efficiently obtained via proportionally allocating the time according to $\{\tau_{\omega}\}$ obtained in (P2.1). We denote the above obtained trajectory as $\{\textbf{q}^0[n]\}_{n=1}^N$.

\subsubsection{$ $\texorpdfstring{$T< T_{\mathrm{tsp}}$}{}$ $}
 \label{tsp2}

In this case, it is infeasible for the UAV to visit all the hovering locations within the given time $T$. To design a feasible trajectory, we specify a disk-shaped region centered at each hovering location $\{\textbf{q}_{\omega}\}$ with the same radius $R$. The main idea here is to find the UAV trajectory and radius $R$ such that the UAV is able to reach each disk region. This problem is equivalent to the traveling salesman problem with neighborhoods (TSPN), for which an efficient suboptimal solution can be found in \cite{zhang2018uav}. For notational convenience, we also denote the above trajectory obtained as $\{\textbf{q}^0[n]\}_{n=1}^N$.

Given the initial UAV trajectory, the transmission scheduling and power allocations of SNs can be optimized. Then the UAV trajectory can be further optimized jointly with the transmission scheduling and power allocations. In the following, we consider two sub-problems of (P1), namely, optimizing SNs' transmission scheduling and power allocations with fixed UAV trajectory, and optimizing UAV trajectory with fixed SNs' scheduling and power allocations. Finally, based on the solutions obtained for these two sub-problems, an iterative algorithm is proposed to alternately optimize these two blocks of variables until they both converge.

\subsection{Transmission scheduling and power allocations optimization with fixed trajectory}

In this subsection, we consider the sub-problem to optimize the SNs' transmission scheduling $\{a_k[n]\}$ and power allocations $\{p_k[n]\}$ with fixed UAV trajectory $\{\textbf{q}[n]\}$. This problem is given by
\begin{subequations}
\begin{align}
\mathrm{(P3)}~~\max_{\{p_k[n]\},\{a_k[n]\},r}~~r      ~~~~~~~~~~~~~~~              \nonumber \\
\label{p3001}
\!\!\!\!\mathrm{s.t.}~~ \frac{1}{N} \sum_{n=1}^N a_k[n]\log_2\left(1+\frac{\kappa_n p_k[n]\gamma_0}{d_k[n]^{\alpha}} \right) \geq r, \ \  k \in \mathcal{K}, \!\!\!\!\! \\
\label{p3002}
\sum_{k=1}^K a_k[n] \leq M, \ \ \forall n, ~~~~~~~~~~~~~~~~~~~~~~~~~~~~~~~ \\
\label{p3003}
a_k[n] \in \{0,1\}, \ \ k \in \mathcal{K}, \forall n, ~~~~~~~~~~~~~~~~~~~~~~~ \\
\label{p3004}
\frac{1}{N}\sum_{n=1}^N p_k[n] \leq \bar{P}, \ \ k \in \mathcal{K},~~~~~~~~~~~~~~~~~~~~~~~~ \\
\label{p3005}
p_k[n] \geq 0, \ \ k \in \mathcal{K},  \forall n.~~~~~~~~~~~~~~~~~~~~~~~~~~~~~
\end{align}
\end{subequations}
Although (P3) is also a mixed-integer non-convex optimization problem, it can be verified that it satisfies the time-sharing condition in \cite{yu2006dual} such that it can be optimally solved using the Lagrange duality method, which is similar to problem (P2). The details for solving (P3) are given in Appendix.

\subsection{Trajectory optimization with fixed transmission scheduling and power allocations}

In this subsection, we consider the other sub-problem~to further optimize the UAV trajectory  $\{\textbf{q}[n]\}$ given the~SNs' transmission scheduling $\{a_k[n]\}$ and power allocations $\{p_k[n]\}$ obtained by solving (P3). The problem is formulated as
\begin{subequations}
\begin{align}
\mathrm{(P4)}~~\max_{\{\textbf{q}[n]\},r}~~r ~~~~~~~~~~~~~\nonumber \\
\label{p31001}
\!\!\!\!\!\!\!\!\!\!\!\!\!\!\!\!\mathrm{s.t.}~~\frac{1}{N}\sum_{n=1}^N \log_2\left(1+\frac{\varepsilon_k[n]}{(z[n]^2+||\textbf{q}[n]-\textbf{d}_k||^2)^{\alpha/2}} \right) \geq r,\!\!\!\!\!\!\!\!\!\!\!\!\!\!\!\!\!\!\!\!\!~ \nonumber\\ k \in \mathcal{K}, \\
\label{p31002}
||\textbf{q}[n+1]-\textbf{q}[n]|| \leq V_{\mathrm{h}}, \ \ n=1,\cdots,N-1,\!\!\! \\
\label{p31003}
\textbf{q}[1]=\textbf{q}_{\mathrm{I}}, \ \ \textbf{q}[N]=\textbf{q}_{\mathrm{F}},~~~~~~~~~~~\!~~~~~~~~~~~~~
\end{align}
\end{subequations}
where $\varepsilon_k[n]\triangleq \kappa_n p_k[n]\gamma_0$. Although problem (P4) is a non-convex optimization problem due to the non-convex constraints in \eqref{p31001}, an efficient suboptimal solution can be found by applying the SCA technique with any given initial UAV trajectory (e.g., $\{\textbf{q}^0[n]\}$ obtained in Section \ref{tspini}). To this end, we need the following result.
\begin{Lemma}
For any given trajectory $\{\textbf{q}^{j}[n]\}$, we have
\begin{align}
\!\!\!\!\!\!r_k[n] \geq r^{lb}_k[n] \nonumber~~~~~~~~~~~~~~~~~~~~~~~~~~~~~~~~~~~~~~~~~~~~~ \\
\triangleq a_k[n] \log_2\left(1+\frac{\varepsilon_k[n]}{(z[n]^2+||\textbf{q}^{j}[n]-\textbf{d}_k||^2)^{\alpha/2}}   \right)   \nonumber\!\!\!\!\!\! \\
-\vartheta^j_k[n]\left(||\textbf{q}[n]-\textbf{d}_k||^2-||\textbf{q}^{j}[n]-\textbf{d}_k||^2 \right),~~
\end{align}
where
\begin{align}
\!\!\!\!\!\!\!\!\!\!\vartheta^{j}_k[n]\triangleq ~~~~~~~~~~~~~~~~~~~~~~~~~~~~~~~~~~~~~~~~~~~~~~~~~~~~ \nonumber \\ \!\!\!\!\!\!\!\!\!\!\!\!\frac{a_k[n](\log_2e)\varepsilon_k[n](\alpha/2)}{\left(z[n]^2+||\textbf{q}^{j}[n]-\textbf{d}_k||^2\right) \left(
\left(z[n]^2+||\textbf{q}^{j}[n]-\textbf{d}_k||^2\right)^{\alpha/2}+\varepsilon_k[n]\right)}. \!\!\!\!\!\!\!\!\!
\end{align}
\end{Lemma}
\begin{IEEEproof}
The proof is similar to that given in \cite{zeng2019accessing}, \cite{zeng2016throughput}, and thus omitted for brevity.
\end{IEEEproof}

As a result, a lower bound of the optimal value to problem (P4) with fixed SNs' transmission scheduling and power allocations can be obtained via solving the following problem
\begin{align}
\mathrm{(P4.1)}~~\max_{\{\textbf{q}[n]\},r}~~r ~~~~~~~~~~~~~~~~~~~~~\nonumber \\
\mathrm{s.t.}~~\frac{1}{N}\sum_{n=1}^N  r^{lb}_k[n] \geq r, \ \ k \in \mathcal{K}, ~~~~\\
\eqref{p31002},~\eqref{p31003}. ~~~~~~~~~~~~~~~~~~~~ \nonumber
\end{align}
Problem (P4.1) can be verified to be convex, which can be efficiently solved by CVX \cite{grant2014cvx}.

Finally, based on the solutions obtained to the above two sub-problems, the overall algorithm to solve problem (P1) is summarized in Algorithm \ref{algo2}. Although the TSP involved in step 2 is an NP-hard problem, it can be efficiently solved with high accuracy by existing algorithms with polynomial complexity. In step 5, the complexity for obtaining the scheduling and power allocation solution is similar to that of Algorithm \ref{algo1}, which is $O((NK_{\mathrm{max}}\Delta_G+K^2)K^2)$. The complexity for updating the trajectory in step 6 is $O(N^{3.5})$ \cite{zhang2018uav}. By denoting the number of iterations needed for convergence as $J$, the total complexity of Algorithm \ref{algo2} is $O(J( (NK_{\mathrm{max}}\Delta_G+K^2)K^2 +N^{3.5}))$.

\begin{algorithm}
\caption{Proposed Algorithm for Solving (P1)\label{algo2}.}
1:~Solve problem (P2) with Algorithm \ref{algo1} to obtain the hovering\\
 \hspace*{0.3cm} locations $\{\textbf{q}^*_{\omega}\}$, scheduling $\{a^*_{k,\omega}\}$, and power allocations\\
 \hspace*{0.3cm} $\{p^*_{k,\omega}\}$, $\omega=1,\cdots,\Omega^*$.\\
2:~Solve the TSP to obtain the permutation order $\boldsymbol{\pi}$ and ~\!\\
 \hspace*{0.3cm} flying time $T_{\mathrm{tsp}}$ to visit all the above hovering locations. \\
3:~Construct the initial trajectory $\{\textbf{q}^0[n]\}$ by comparing $T$ \\
 \hspace*{0.3cm} and $T_{\mathrm{tsp}}$ according to Section \ref{tspini}. Let $j$$=$$0$.\\
4:~\textbf{Repeat} \\
5:~~~~For given trajectory $\{\textbf{q}^j[n]\}$, obtain the optimal\\
\hspace*{0.63cm} transmission scheduling $\{a^{j+1}_k[n]\}$ and power allocations \\
\hspace*{0.6cm} $\{p^{j+1}_k[n]\}$  based on the algorithm given in Appendix.\\
6:~~~~~\!~For given $\{a^{j+1}_k[n]\}$ and $\{p^{j+1}_k[n]\}$, update the trajectory
\hspace*{0.6cm} $\{\textbf{q}^{j+1}[n]\}$ via solving (P4.1). \\
7:~~~~Update $j=j+1$. \\
8:~\textbf{Until} the objective value $r$ converges within a given \\
\hspace*{0.3cm} accuracy.
\end{algorithm}

\section{Simulation Results}
\label{simul}

\begin{figure*}[!htb]
\vspace{-4.1cm}
\minipage{0.32\textwidth}
  \hspace*{-1.9cm}\includegraphics[width=1.65\linewidth]{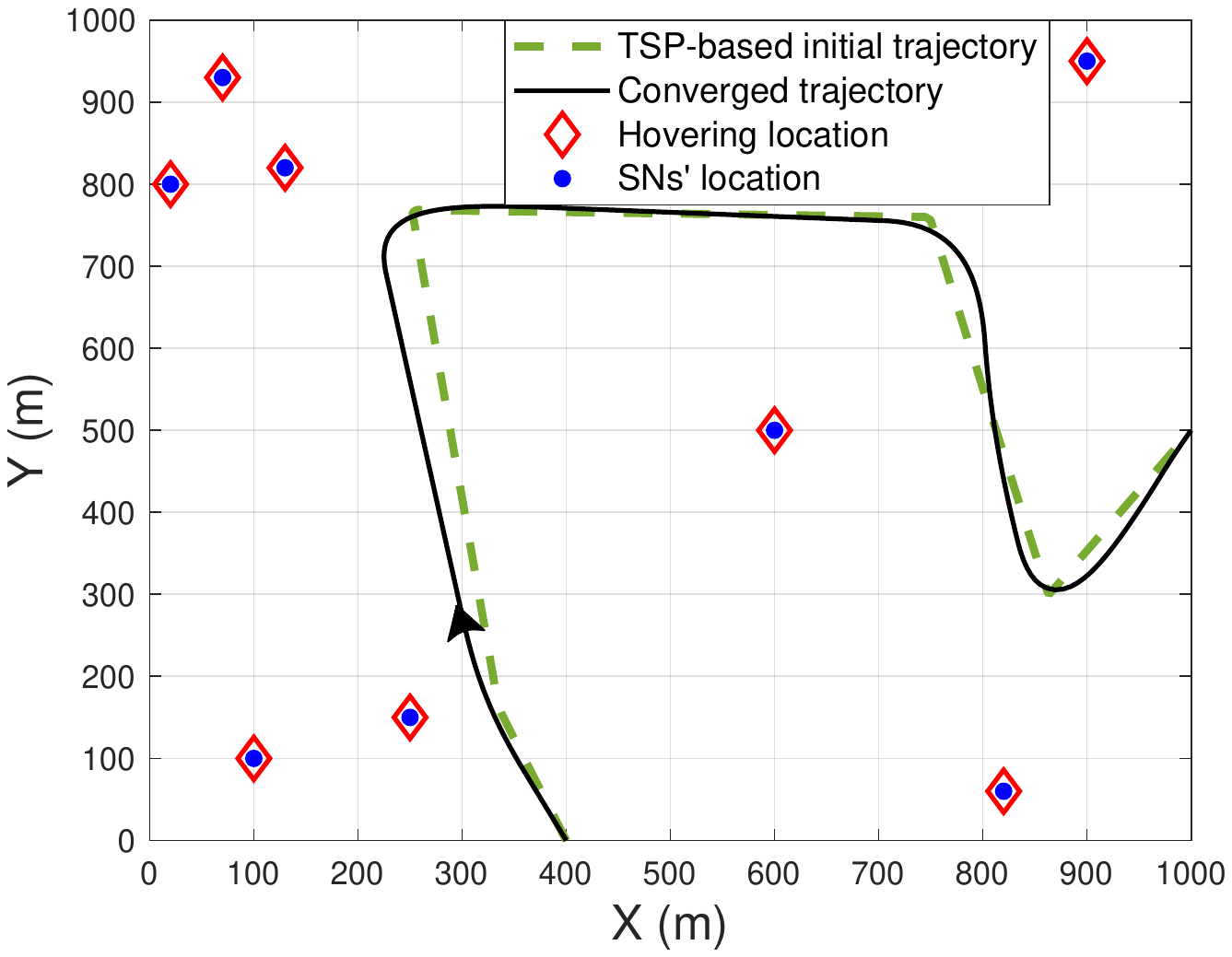}
\vspace{-4.4cm}
  \subcaption{\label{mrc100}MRC scheme with $M=12$.}
\endminipage\hfill
\minipage{0.32\textwidth}
\hspace*{-1.8cm}\includegraphics[width=1.65\linewidth]{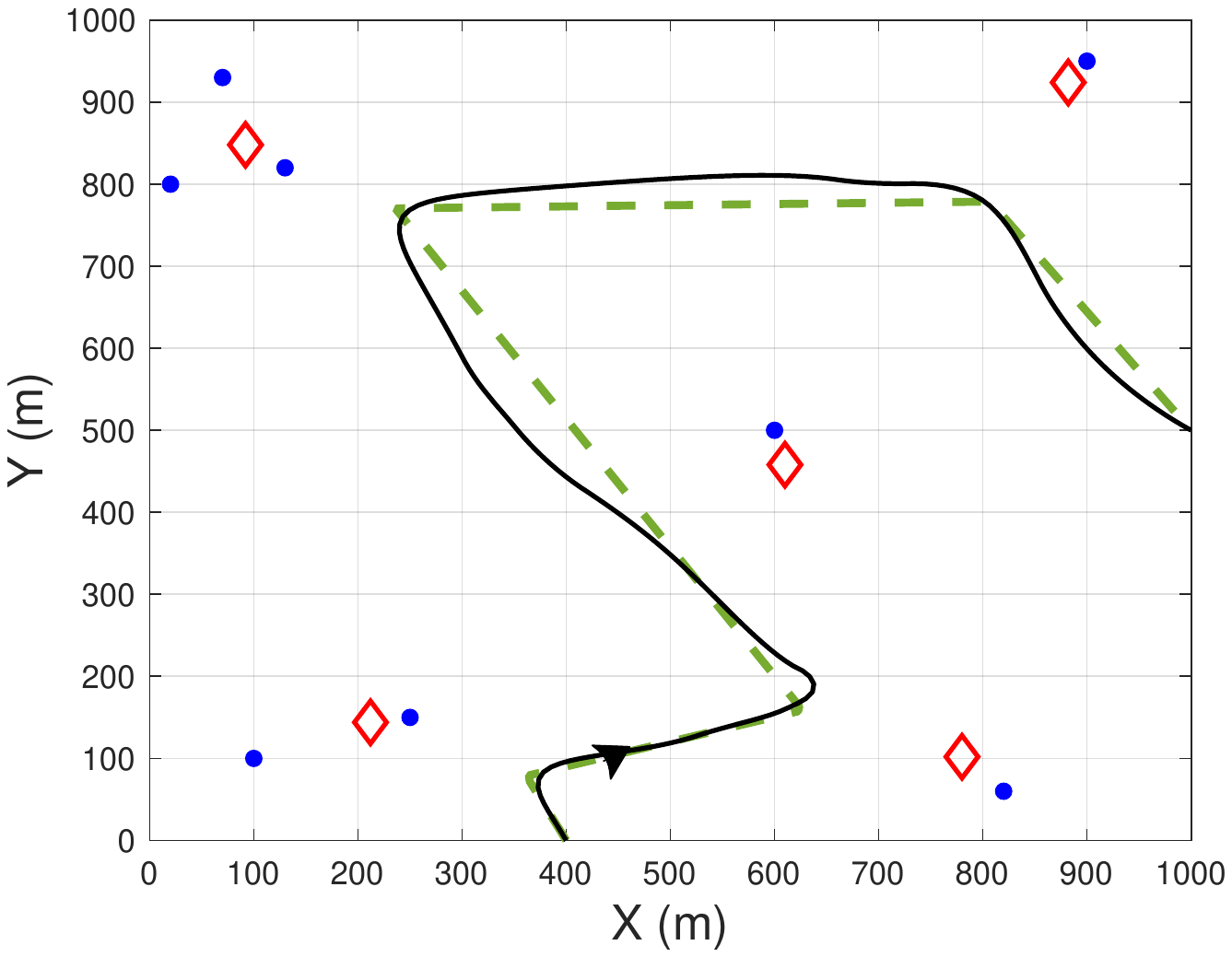}
\vspace{-4.4cm}
  \subcaption{\label{p12_100}Proposed scheme with $M=12$.}
\endminipage\hfill
\minipage{0.32\textwidth}%
 \hspace*{-1.7cm} \includegraphics[width=1.65\linewidth]{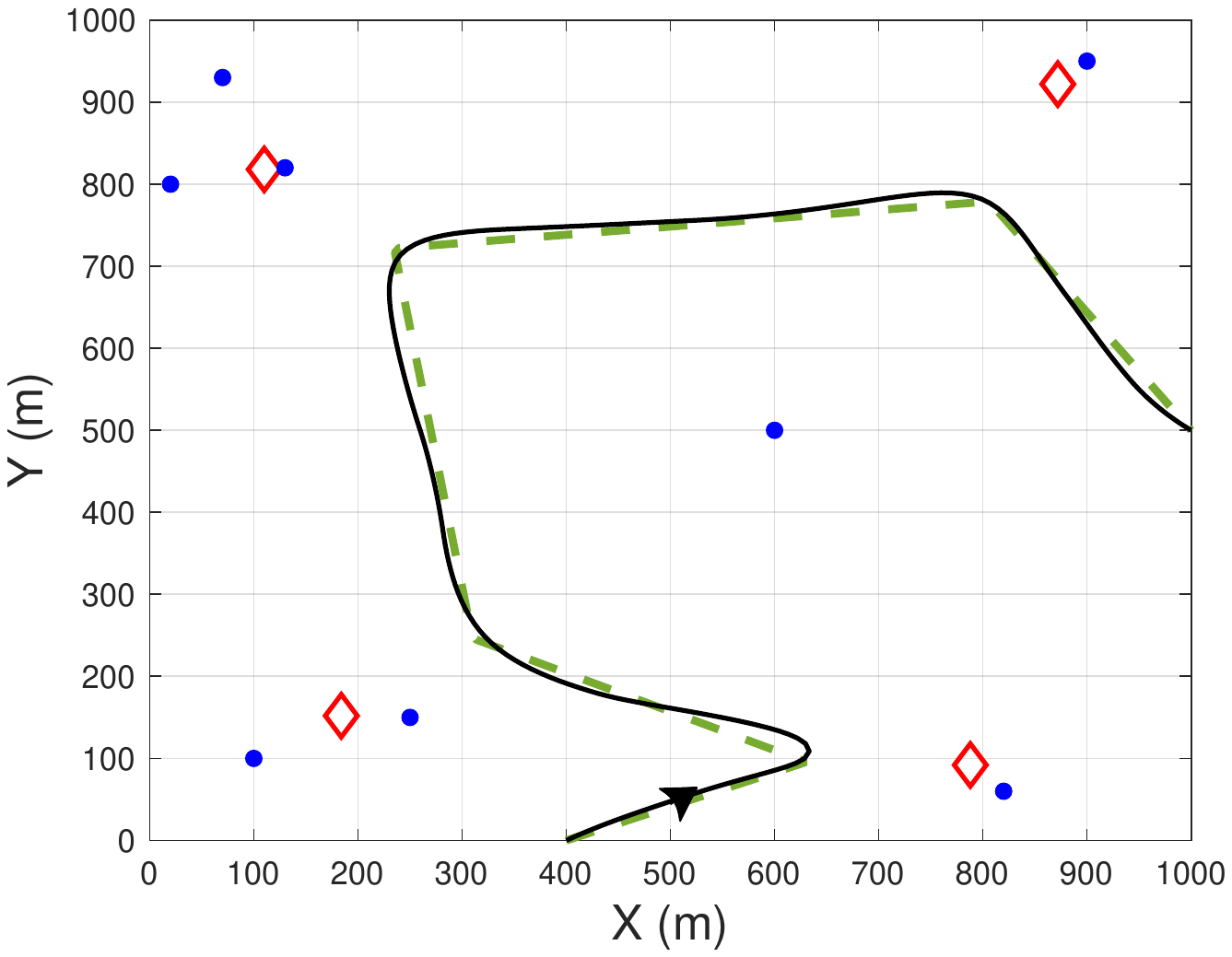}
\vspace{-4.4cm}
  \subcaption{\label{p20_100}Proposed scheme with $M=20$.}
\endminipage\hfill
\vspace{-0.1cm}
\caption{\label{100} Hovering locations and trajectories of different schemes under $T=100$ s. }
\vspace{-0.5cm}
\end{figure*}

In this section, numerical results are provided to evaluate the performance of the proposed design for multi-antenna UAV data harvesting. We consider a WSN with $K=8$ SNs that are distributed in a square area with side length equal to 1000 m. The simulation results are based on one realization of SNs' locations, as shown in Fig. \ref{mrc100}. The channel power gain at the reference distance $d_0=1$ m is set as $\beta_0=-60$ dB. The total available bandwidth is $B=0.1$ MHz and the noise power spectrum density is $N_0=-154$ dBm/Hz such that the noise power $\sigma^2=B N_0=-104$ dBm. The average power limit at all SNs is set as $\bar{P}=0.01$ W.
The path loss exponent is set as $\alpha=2$ and the Rician factor is $G=0.94$ \cite{1469712}. The UAV flying altitude is set as $H_{\mathrm{min}}=130$ m, and the initial and final locations are $\textbf{q}_{\mathrm{I}}=[400,0]^T$ m and $\textbf{q}_{\mathrm{F}}=[1000,500]^T$~m, respectively. Furthermore, the maximum UAV speed is $v_{\mathrm{h}}=20$ m/s, and the length of time slot is set as $\delta=0.5$ s. We assume that a uniform rectangular array (URA) with 4 antennas per row is mounted on the UAV, where the adjacent antennas are equally separated both horizontally and vertically. The number of antennas at the UAV is $M=4$, $12$ or $20$.


For comparison, two benchmark schemes are considered: 1) MRC receive beamforming scheme, where at most one SN can be scheduled for transmission in each time slot, i.e., $K_n=1$ and $\kappa_n=M$, $\forall n$; 2) single-antenna scheme, where only one antenna is mounted on the UAV, i.e., $M=1$. Specifically, for the single-antenna scheme, only one SN can transmit in each time slot and the achievable rate can be approximated as \eqref{R2} with $\kappa_n=1$, $\forall n$.

\begin{figure}
\vspace{-4.4cm}
\hspace*{-0.7cm}\includegraphics[width=4in]{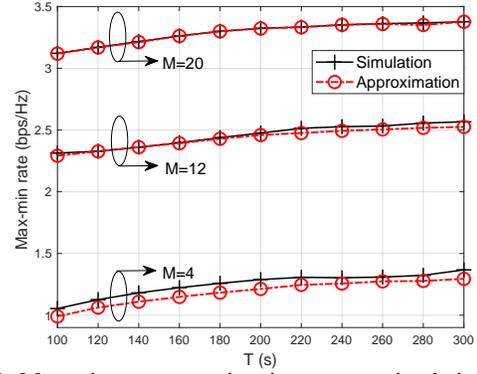}
\vspace{-4.6cm}
\caption{\label{approximation}Max-min rate approximations versus simulation results.}
\vspace{-0.6cm}
\end{figure}

\begin{figure*}[!htb]
\vspace{-3.7cm}
\minipage{0.32\textwidth}
  \hspace*{-1.9cm}\includegraphics[width=1.65\linewidth]{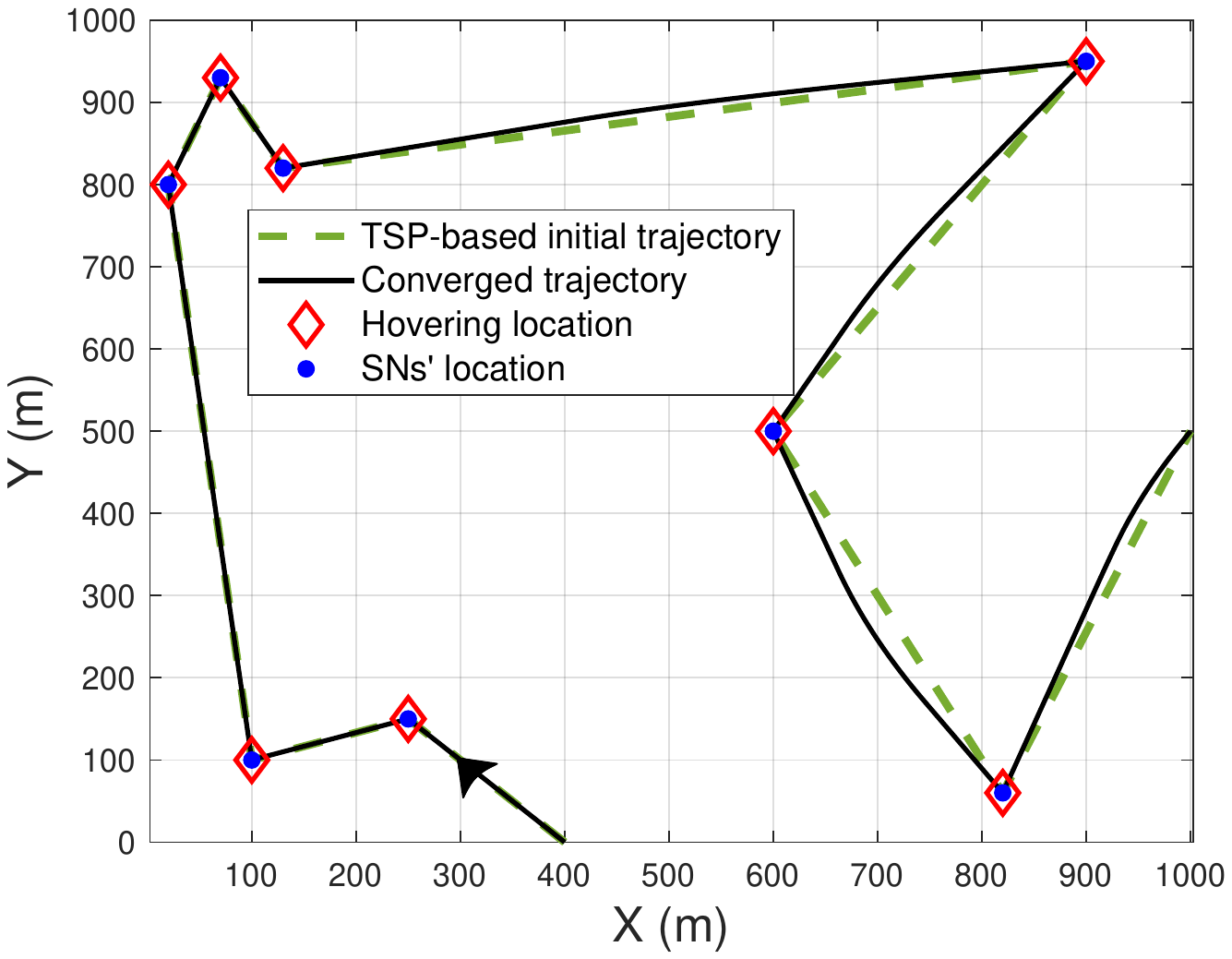}
\vspace{-4.4cm}
  \subcaption{\label{mrc200}MRC scheme with $M=12$.}
\endminipage\hfill
\minipage{0.32\textwidth}
\hspace*{-1.8cm}\includegraphics[width=1.65\linewidth]{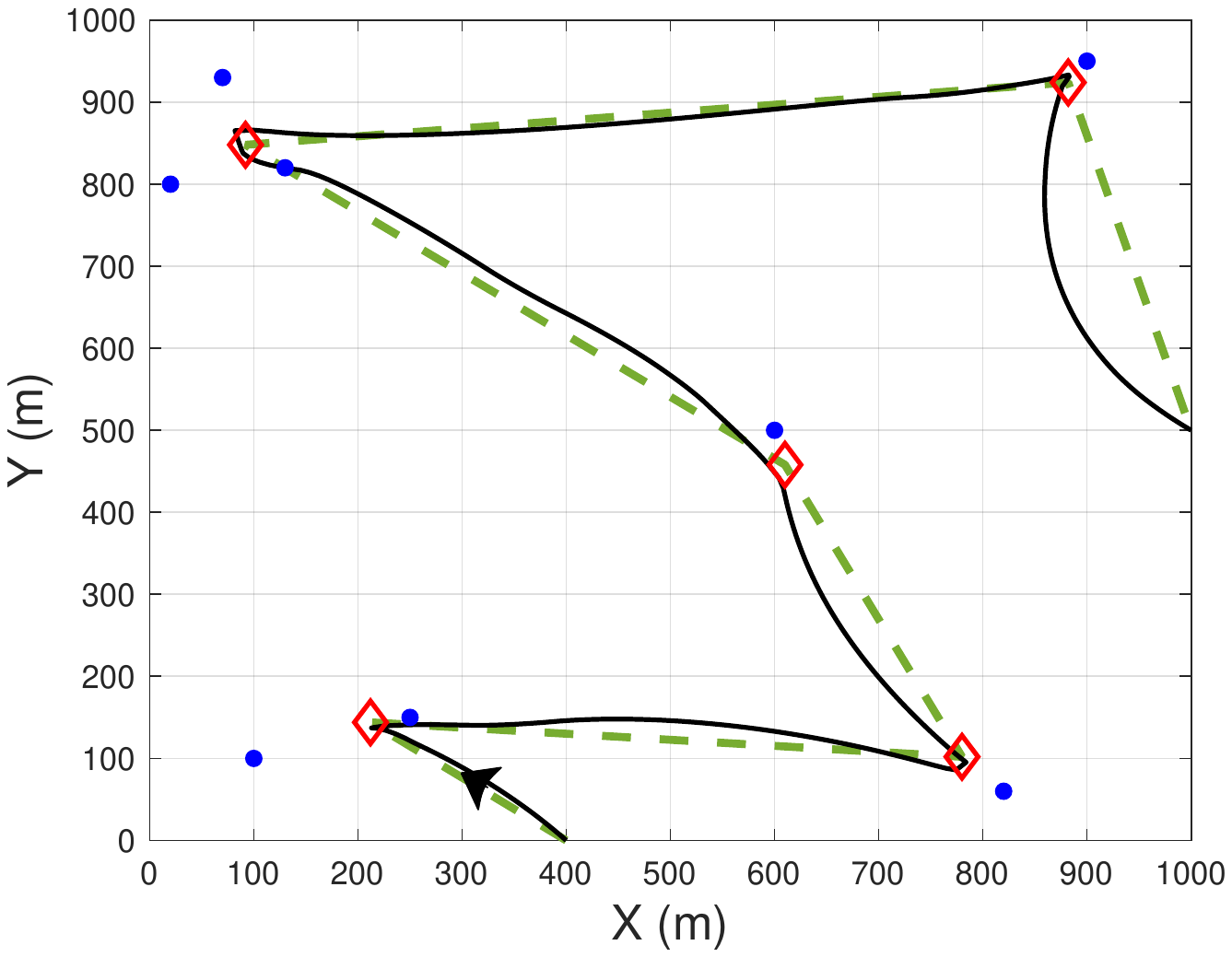}
\vspace{-4.4cm}
  \subcaption{\label{p12_200}Proposed scheme with $M=12$.}
\endminipage\hfill
\minipage{0.32\textwidth}%
 \hspace*{-1.7cm} \includegraphics[width=1.65\linewidth]{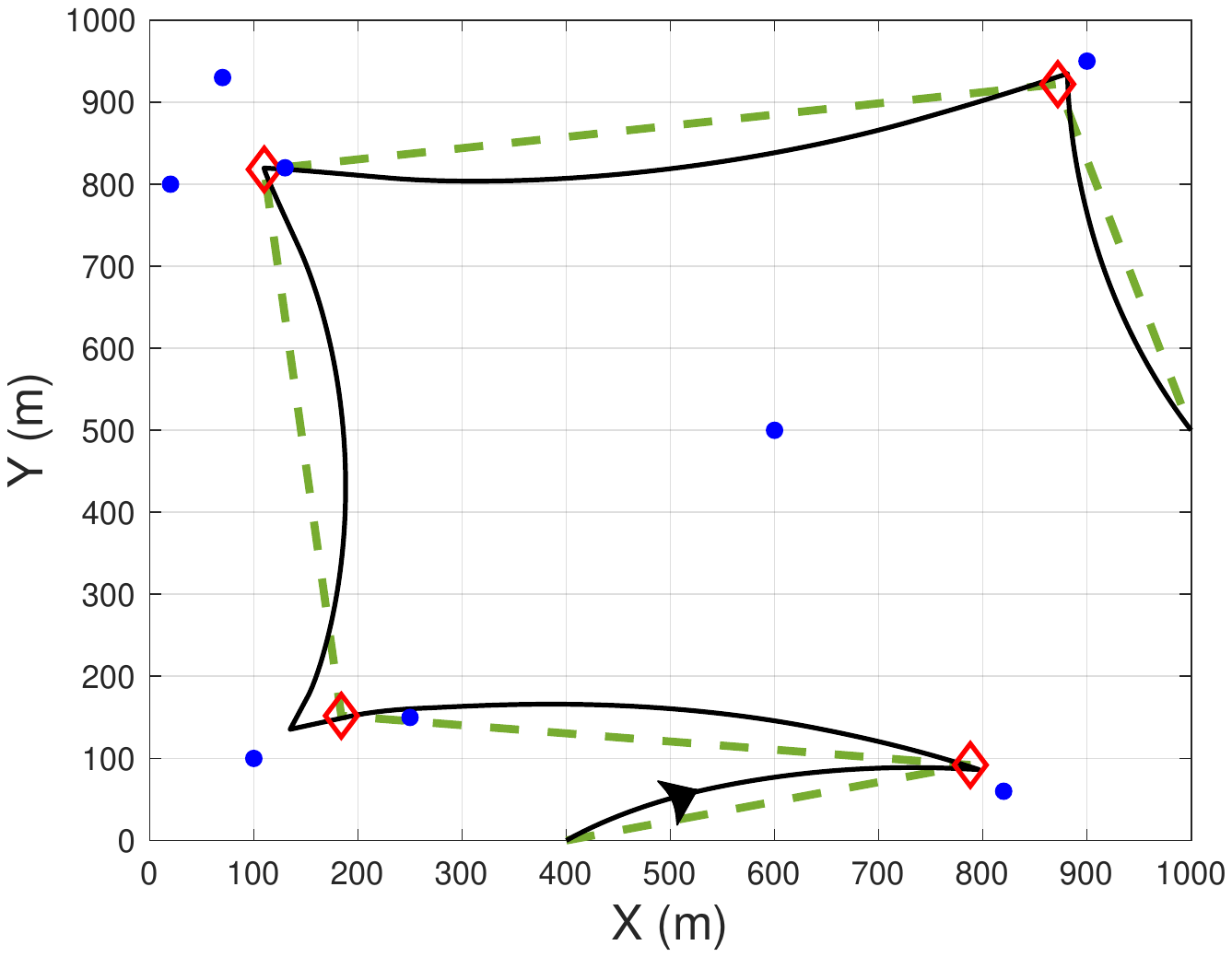}
\vspace{-4.4cm}
  \subcaption{\label{p20_200}Proposed scheme with $M=20$.}
\endminipage\hfill
\vspace{-0.1cm}
\caption{\label{200} Hovering locations and trajectories of different schemes under $T=200$ s. }
\vspace{-0.5cm}
\end{figure*}

First, in order to evaluate the accuracy of the proposed rate approximation given in \eqref{R2}, Fig. \ref{approximation} shows the achievable max-min rate for all SNs computed via numerical simulations compared with the closed-form rate approximation in \eqref{R2} with different number of antennas, where the rate is averaged over $10^3$ random channel realizations at each UAV location. It is observed that the approximation \eqref{R2} achieves good accuracy, especially for large number of antennas, which is also in accordance with \cite{liu2019comp}. In particular, there exists a small gap between the numerical simulation and the approximation for small number of antennas, e.g., $M=4$, while such gap can be practically ignored for sufficiently large number of antennas, e.g., $M=12,20$.

Next, we evaluate the performance of the proposed design for the minimum-rate maximization problem. Besides the optimal hovering locations obtained by solving (P2) without considering the UAV speed constraint, in Figs. \ref{100} and \ref{200}, the converged UAV trajectories obtained by solving (P1) with the speed and initial/final location constraints are also shown for different schemes under $T=100$ s and 200 s, respectively. First, it is observed that for the MRC scheme in the case of $T=100$ s as shown in Fig. \ref{mrc100}, there are 8 UAV hovering locations, each above a different SN for receiving its data. When the number of antennas for the MRC scheme is $M=4$ or $M=20$, the optimal hovering locations of the UAV are the same as that for $M=12$ and thus are not shown for brevity. The max-min rates achieved by the MRC scheme after solving (P2) with $M=4$, 12, and 20 are 1.04 bps/Hz, 1.27 bps/Hz, and 1.37 bps/Hz, respectively; thus, the rate gain with the increasing number of antennas is only marginal. The reason is that under the MRC scheme, only the beamforming gain is achieved and the UAV needs to receive data for each SN over orthogonal time slots, thus resulting in low spectrum efficiency. As for the single-antenna scheme, it is obvious that at the optimal solution to (P2), the UAV should hover above each SN with the same duration, similar to the case of the MRC scheme. Thus, the optimal hovering locations and converged trajectories of this scheme are omitted for brevity. The max-min rate achieved by the single-antenna scheme after solving (P2) is 0.74 bps/Hz, which is lower than those obtained by the MRC scheme, as expected.

The optimal hovering locations obtained by the proposed scheme for solving (P2) are shown in Figs. \ref{p12_100} and \ref{p20_100}, respectively, for the case of $T=100$ s. It is observed that the number of UAV hovering locations is less than that in the two benchmark schemes, and it decreases with the increasing number of antennas, which is 5 for $M=12$ and 4 for $M=20$. Specifically, 3 SNs in the upper-left corner or 2 SNs in the lower-left corner need to be served by the UAV at one single hovering location under the proposed scheme, rather than one hovering location for each SN under the MRC/single-antenna scheme. Moreover, for the proposed scheme, different from the case with $M=12$ where one hovering location is needed to cover the SN around the center of the area of interest, it is no longer needed when the number of antennas increases to $M=20$. The max-min rate achieved with $M=12$ and $M=20$ under the proposed scheme after solving (P2) is 2.57 bps/Hz and 3.41 bps/Hz, respectively, which are significantly higher than those of the MRC scheme or the single-antenna scheme. The reason is that different from the two benchmark schemes, where the UAV only serves one SN in each time slot, the UAV is able to serve multiple SNs simultaneously under the proposed scheme by exploiting the spatial multiplexing gain with ZF receive beamforming; as a result, fewer hovering locations are needed for the UAV. Accordingly, the flying distance of the UAV can also be reduced after solving (P1) with the speed constraint, such that the UAV has more time to stay at the hovering locations to achieve higher rate, as shown next. Note that similar results can be observed for the case of $T=200$ s in Fig. \ref{200}.

\begin{figure*}
\vspace{-4.1cm}
\begin{minipage}{0.6\linewidth}
\vspace{-0.62cm}
\hspace*{-0.6cm}\includegraphics[width=4in]{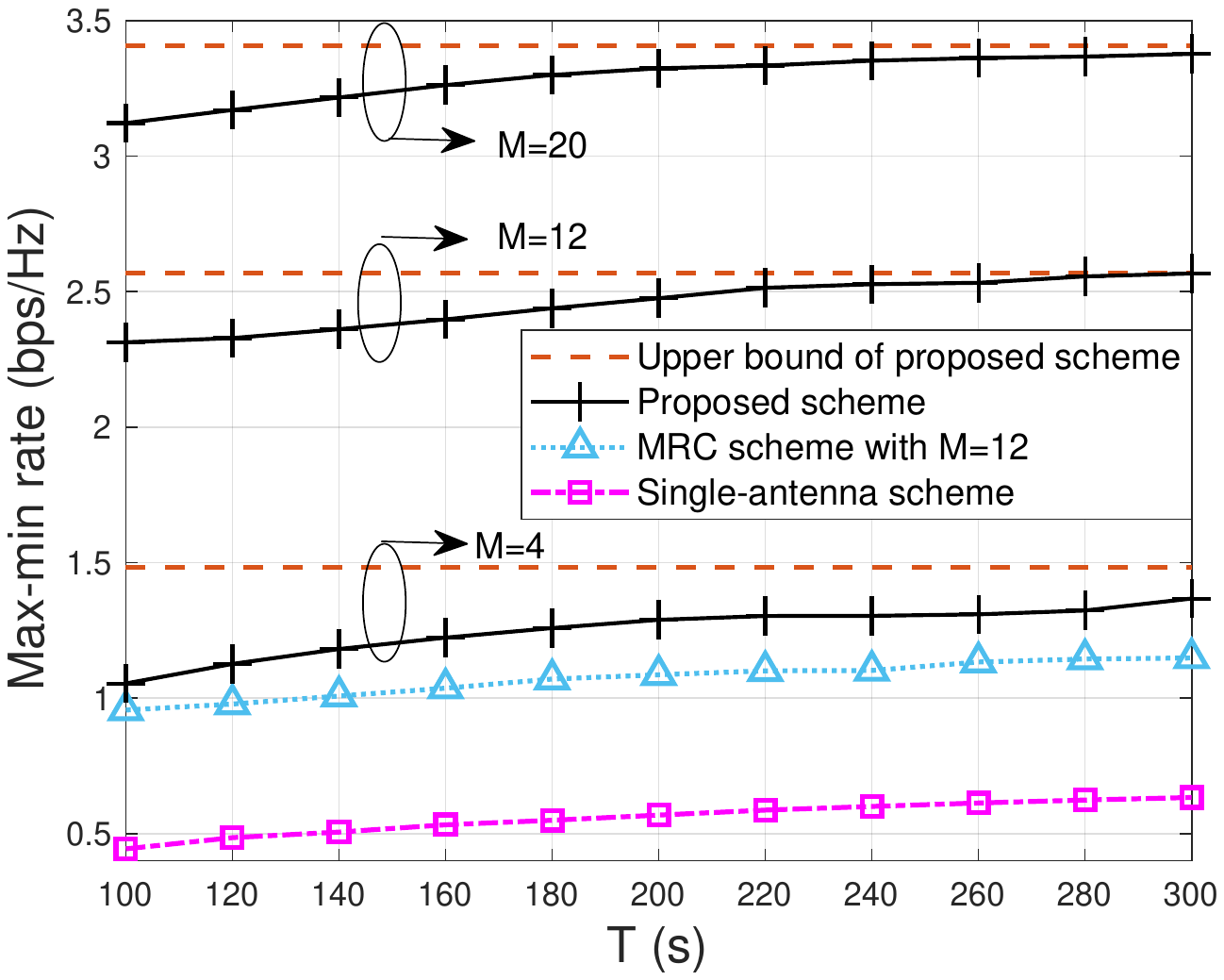}
\vspace{-4.2cm}
\caption{\label{rate_comp}Max-min rate comparison between different schemes.~~~~~~~~~~~~~~ }
\end{minipage}
\vspace{0.2cm}
\hspace*{-2.6cm}\begin{minipage}{0.6\linewidth}
\vspace{-0.25cm}
\hspace*{0.2cm}\includegraphics[width=4in]{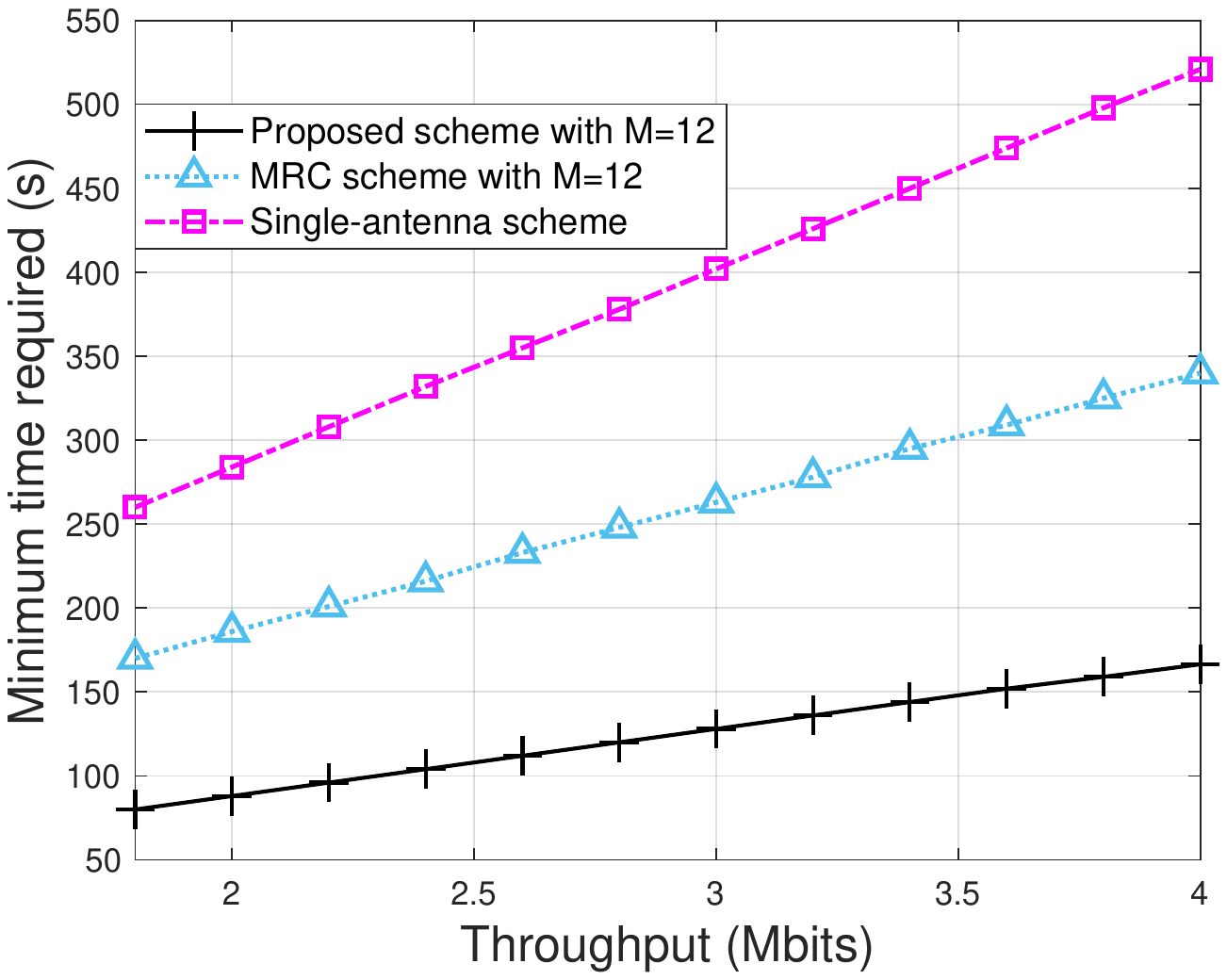}
\vspace{-4.2cm}
\caption{\label{thrpt_time}Minimum flight time required versus common throughput \\ requirement. }
\end{minipage}
\vspace{-0.2cm}
\vspace{-0.4cm}
\end{figure*}

In addition, the initial and converged trajectories of the UAV for solving (P1) with the speed constraint under $T=100$ s and $T=200$ s are also shown in Figs. \ref{100} and \ref{200}, respectively. The minimum time required for the UAV to visit all hovering locations for the MRC scheme and that for the proposed scheme with $M=12$ and $M=20$ is 181 s, 154 s, and 144 s, respectively. Hence, in the initial trajectory design, when the given time $T=100$ s is insufficient for the UAV to visit all hovering locations, the UAV tries to fly as close as possible to each hovering location to shorten the link distance with the SNs, as shown in Fig. \ref{100}. While when the given time increases to $T=200$ s, the UAV can successively visit all hovering locations, as shown in Fig. \ref{200}. Besides, with shorter flying distance and thus time by the proposed scheme as shown in Figs. \ref{p12_200} and \ref{p20_200}, the UAV has more time to stay at the hovering locations to achieve higher max-min rate, which is 2.48 bps/Hz for $M=12$ and 3.32 bps/Hz for $M=20$, as compared to 1.09 bps/Hz for the MRC scheme shown in Fig. \ref{mrc200}.

The max-min rate achieved by solving (P1) under different $T$ by the considered schemes is compared in Fig. \ref{rate_comp}, where that achieved by solving (P2) without the speed constraint is also shown as the rate upper bound. It is observed that for all the three schemes considered, the max-min rate gradually converges to the rate upper bound with the increase of $T$. This is expected since with a longer flight duration, the UAV has more time to stay at the hovering locations obtained in (P2) and thus the rate achieved when it is flying becomes more negligible. Furthermore, the achievable rate of the proposed scheme is significantly improved with the increasing number of antennas at the UAV and also greatly outperforms those of the benchmark schemes.

Fig. \ref{thrpt_time} compares the minimum flight time required for the UAV by different schemes versus the common (minimum) throughput requirement of all SNs. Since the proposed scheme with $M=20$ significantly outperforms other schemes as shown in Fig. \ref{rate_comp}, we only consider the proposed scheme with $M=12$, MRC scheme with $M=12$, and the single-antenna scheme for ease of comparison. The minimum time required for different throughput requirements can be found by solving (P1) with an additional bisection search. It is observed in Fig. \ref{thrpt_time} that the proposed scheme always needs the least flight time to meet the throughput requirement of the SNs, compared to the two benchmark schemes. On the other hand, with less time to meet the throughput requirement, the energy consumption of the UAV can also be saved, as shown next.

\begin{figure}
\vspace{-4.2cm}
\hspace*{-0.7cm}\includegraphics[width=4in]{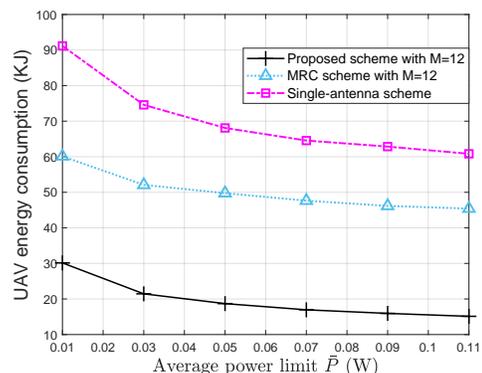}
\vspace{-4.6cm}
\caption{\label{energy}Trade-off between average transmit power limit of SNs and \\ energy consumption of UAV.}
\vspace{-0.6cm}
\end{figure}

Finally, to show the benefits brought by multi-antenna UAV data harvesting in terms of energy saving, we consider the trade-off between the average transmit power limit of SNs and the energy consumption of the UAV. In general, the energy consumption of the UAV consists of two parts for our considered problem. The first part is the propulsion energy, while the second part is the communication-related energy, which is much smaller than the former for practical UAVs and thus is ignored for simplicity. Specifically, based on \cite{zeng2019energy}, the propulsion power of the UAV can be modelled as
\begin{align}
\!\!\!\!\!\!\!\!\!\!\!\!\!\!\!\!\!\!\!\!\!\!\!\!\!\!\!\!\!\!\!\!P_{\mathrm{h}}[n]=P_0\left(1+\frac{3v_{\mathrm{h}}[n]^2}{U_{\mathrm{tip}}^2}  \right)+P_i\left(\sqrt{1+\frac{v_{\mathrm{h}}[n]^4}{4v_0^4}}-\frac{v_{\mathrm{h}}[n]^2}{2v_0^2} \right)^{1/2} \nonumber \!\!\!\!\!\!\!\!\!\!\!\!\!\!\!\!\!\!\!\!\!\!\!\!\!\!\!\!\!\!\!\!\!\!\!\!\\
+\frac{1}{2}d_1 \varrho s A v_{\mathrm{h}}[n]^3,
\end{align}
where $v_{\mathrm{h}}[n]=||(\textbf{q}[n+1]-\textbf{q}[n])||/\delta$ is the horizontal speed of the UAV, $P_0=  79.8563$ and $P_i= 88.6279$ are two constants, $U_{\mathrm{tip}}=120$ m/s represents the tip speed of the rotor blade, $v_0=4.03$  is the mean rotor induced velocity in hover, $d_1=0.6$ and $s=0.05$ are the fuselage drag ratio and rotor solidity, respectively, $\varrho=1.225$ $\mathrm{kg}/\mathrm{m}^3$ and $A=0.503$ $\mathrm{m}^2$ denote the air density and rotor disc area, respectively. Fig. \ref{energy} shows the trade-off between the average transmit power limit of SNs and the (propulsion) energy consumption of the UAV for different schemes under the same throughput requirement of 4 Mbits for all SNs. It is observed that for all the three schemes, as the average transmit power $\bar{P}$ of the SNs increases, the energy consumption of the UAV decreases, which demonstrates the energy trade-off between SNs and the UAV \cite{8316986}, and such a trade-off is more evident for smaller $\bar{P}$. This is expected since with higher transmit power of SNs, the UAV generally needs less flight time to meet their throughput requirement, thus leading to lower energy consumption. On the other hand, it is observed that the proposed scheme always outperforms both benchmark schemes in terms of the SNs-UAV energy consumption trade-off, as expected.

\section{Conclusion}
\label{conc}

This paper studies a UAV-enabled WSN, where a multi-antenna UAV is employed to collect data from a group of single-antenna SNs. Our aim is to maximize the minimum rate of all SNs via jointly optimizing the transmission scheduling and power allocations of SNs as well as the UAV trajectory, subject to practical SNs' transmit power and UAV speed constraints. To tackle this challenging non-convex optimization problem, we first consider its relaxed problem by ignoring the UAV maximum speed constraint and derive the optimal solution to it. Based on this solution, we further propose a suboptimal solution to the general problem with the UAV speed and initial/final location constraints. Numerical results are provided to show significant performance gains by the proposed scheme, as compared to benchmark schemes, in terms of both spectrum efficiency and SNs-UAV energy consumption trade-off.

\section*{Appendix}
\label{Appendix}

\section*{Optimal Solution to (P3) }

 In this appendix, we derive the optimal solution to problem (P3). Since the problem can be optimally solved by the Lagrange duality method similar to (P2), we first express the partial Lagrangian of (P3) as
\begin{align}
\label{l2}
\mathcal{L}_2(\{p_k[n]\},\{a_k[n]\},r,\boldsymbol{\nu},\boldsymbol{\phi})=     ~~~~~~~~~~~~~~~~~~~~~~~~~~~~~~~      \nonumber \\
\left(1-\sum_{k=1}^K \nu_k \right)r+\sum_{k=1}^K \nu_k \bar{r}_k+\sum_{k=1}^K\phi_k\left(N \bar{P}-\sum_{n=1}^N p_k[n] \right),
\end{align}
where $\boldsymbol{\nu}\triangleq \{\nu_k  \}$ and $\boldsymbol{\phi} \triangleq  \{\phi_k\}$ are the dual variables associated with the constraints \eqref{p3001} and \eqref{p3004}, respectively. The Lagrange dual function of (P3) is thus given by
\begin{align}
\label{g2}
 \!\!\!\!\!\!g_2(\boldsymbol{\nu},\boldsymbol{\phi})=\nonumber ~~~~~~~~~~~~~~~~~~~~~~~~~~~~~~~~~~~~~~~~~~~~~~~~~\\
\!\!\!\left\{
              \begin{aligned}
              \max_{\substack{\{p_k[n]\}, \{a_k[n]\},r}} \mathcal{L}_2\left(\{p_k[n]\},\{a_k[n]\},r,\boldsymbol{\nu},\boldsymbol{\phi}\right)~~~~~~ \\
              \mathrm{s.t.}~~~~~~~~\eqref{p3002},~\eqref{p3003},~\eqref{p3005},~~~~~~~~~~~~~~~~~
                            \end{aligned}
           \right. \!\!\!\!\!\!\!\!
\end{align}

For \eqref{g2} to be bounded, we should have $1-\sum_{k=1}^K \nu_k =0$. Therefore, the dual problem of (P3) is given by
\begin{subequations}
\begin{align}
\mathrm{(D3)}~~ \min_{\boldsymbol{\nu},\boldsymbol{\phi}} ~~~g_2(\boldsymbol{\nu},\boldsymbol{\phi})  ~~~~~~~~~~~\nonumber \\
\label{d3001}
\mathrm{s.t.}~~~1-\sum_{k=1}^K\nu_k=0,~~~~~~~~~~~ \\
\label{d3002}
\nu_k\geq 0, \ \ \forall k \in \mathcal{K},~~~\!~~~~~~ \\
\label{d3003}
\phi_k \geq 0, \ \ \forall k \in \mathcal{K}. ~~~~~~~~~~\!\!
\end{align}
\end{subequations}

Let the feasible set of $\boldsymbol{\nu}$ and $\boldsymbol{\phi}$ characterized by the constraints in \eqref{d3001}-\eqref{d3003} as $\chi_2$. In the following, we solve (P3) based on the similar procedures as for (P2).

\subsection{Obtaining $g_2($\texorpdfstring{$\boldsymbol{\nu}$}{}$,~\!\!$\texorpdfstring{$\boldsymbol{\phi}$}{}$)$ by solving problem \eqref{g2} for given $($\texorpdfstring{$\boldsymbol{\nu}$}{}$,~\!\!$\texorpdfstring{$\boldsymbol{\phi}$}{}$)$ $\in \mathcal{X}^2$}

Similar to \eqref{waterp} and \eqref{waterr}, the optimal transmit power and achievable rate over time slots are respectively given by
\begin{align}
\label{equati1}
p_k^*[n]=a_k[n] \left[ \frac{\nu_k}{N \phi_k \log(2)}-\frac{d_k[n]^{\alpha}}{\kappa_n\gamma_0} \right]^+,~~~~~ \\
\label{equati2}
r_k^*[n]=a_k[n]\left[ \log_2\left( \frac{\nu_k\kappa_n\gamma_0}{N \phi_k \log(2)d_k[n]^{\alpha}}  \right)  \right]^+.
\end{align}

Then, in \eqref{g2} we have
\begin{align}
 \mathcal{L}_2\left(\{p_k[n]\},\{a_k[n]\},r,\boldsymbol{\nu},\boldsymbol{\phi}\right) \nonumber~~~~~~~~~~~~~~~~~~~~~~~~~~~ \\
\!\!\!\!\!\!\!\!\!\!\!   =\sum_{k=1}^K\nu_k \left( \frac{1}{N}  \sum_{n=1}^N r_k^*[n]  \right)+\sum_{k=1}^K \phi_k\left(N \bar{P}-\sum_{n=1}^N p_k[n]^* \right)  \!\!\! \!\!\! \!\!\!\!\! \nonumber \\
 =\sum_{n=1}^{N} \left( \sum_{k=1}^K a_k[n]\varphi_k[n]  \right)     +N \bar{P}\sum_{k=1}^K \phi_k,~~~~~~~~~~~~~~
\end{align}
where
\begin{align}
\varphi_k[n] \triangleq\frac{\nu_k}{N}\left[\log_2\left( \frac{\nu_k\kappa_n\gamma_0}{N\phi_k \log(2)d_k[n]^{\alpha}} \right) \right]^+ \nonumber \\
-\phi_k\left[ \frac{\nu_k}{N \phi_k \log(2)}-\frac{d_k[n]^{\alpha}}{\kappa_n\gamma_0} \right]^+.
\end{align}
Thus, problem \eqref{g2} is recast to
\begin{align}
\label{g22}
\max_{\{a_k[n]\},\kappa_n}~~\sum_{n=1}^{N} \left( \sum_{k=1}^K a_k[n]\varphi_k[n]  \right)     +N \bar{P}\sum_{k=1}^K \phi_k \\
\mathrm{s.t.}  ~~~~\!~~~\eqref{p3002},~\eqref{p3003}.~ \nonumber~~~~~~~~~~~~~~~~~~~~~~~~~~~
\end{align}

Note that problem \eqref{g22} can be decoupled into $N$ sub-problems, each corresponding to one time slot. For each sub-problem, the optimal scheduling can be easily found via an exhaustive search over all possible values of $\kappa_n$, similar to \eqref{gg2}. However, different from problem \eqref{gg1} where each sub-problem is the same across different time slots, each sub-problem in problem \eqref{g22} is different and thus needs to be solved separately.

\subsection{Finding optimal $ $\texorpdfstring{$\boldsymbol{\nu}$}{}$ $ and $ $\texorpdfstring{$\boldsymbol{\phi}$}{}$ $ to solve (D3) }

Similar to Section \ref{timeshar2}, we can use the ellipsoid method to obtain optimal $\boldsymbol{\nu}$ and $\boldsymbol{\phi}$. The subgradients with respect to $(\boldsymbol{\nu},\boldsymbol{\phi})$ are given by
\begin{align}
\!\Delta \nu_k=\frac{1}{N} \sum_{n=1}^N r_k[n]^*,~~~~\!~  \\
\Delta  \phi_k=N \bar{P}-\sum_{n=1}^N p_k[n]^*,
\end{align}
where $r_k[n]^*$ and $p_k[n]^*$ are given in \eqref{equati2} and \eqref{equati1}, respectively.

Finally, since the obtained $\{a_k[n]\}$ for each sub-problem in \eqref{g22} is unique over different time slots, there is no need to further construct the optimal solution for (P3) by time-sharing as for (P2). The rest of the algorithm for solving (P3) is similar to Algorithm \ref{algo1} for (P2), and thus is omitted for brevity.

\ifCLASSOPTIONcaptionsoff
  \newpage
\fi

\bibliographystyle{IEEEtran}

\bibliography{refer}

\end{document}